\documentstyle[twocolumn,pre,aps,eqsecnum]{revtex}
\include{psfig}

\begin{document}

\draft

\preprint{FSU-SCRI-98-83}

\title{Kinetic Ising model in an oscillating field:
       Avrami theory for the hysteretic response and finite-size scaling
       for the dynamic phase transition}

\date{\today}

\author{S.~W. Sides,$^{1 2 3 a}$
        P.~A. Rikvold,$^{1 2 3 b}$
    and M.~A. Novotny$^{2 c}$}

\address{$^{1}$Center for Materials Research
and Technology and Department of Physics, \\
Florida State University, Tallahassee, Florida 32306-4350 \\
$^{2}$Supercomputer Computations Research Institute, \\
Florida State University, Tallahassee, Florida 32306-4130 \\
$^{3}$Colorado Center for Chaos and Complexity, \\
University of Colorado, Boulder, Colorado 80309-0216 \\
}

\maketitle

\begin{abstract}
Hysteresis is studied for a two-dimensional,
spin-$1/2$, nearest-neighbor, kinetic Ising ferromagnet in a 
sinusoidally oscillating 
field, using Monte Carlo simulations and analytical theory.
Attention is focused on large systems and 
moderately strong field amplitudes at a temperature below $T_{c}$.
In this parameter regime, the magnetization switches through
random nucleation and subsequent growth
of {\it many} droplets of spins aligned with the applied field. Using a
time-dependent extension of the Kolmogorov-Johnson-Mehl-Avrami (KJMA)
theory of metastable decay,
we analyze the statistical properties of the hysteresis-loop area
and the correlation between the magnetization and the field.
This analysis enables us to accurately predict the results of extensive
Monte Carlo simulations.
The average loop area exhibits an extremely slow approach to an
asymptotic, logarithmic dependence on the product of the amplitude
and the field frequency.
This may explain the inconsistent exponent estimates reported
in previous attempts to fit experimental and numerical data
for the low-frequency behavior of this quantity to a power law.
At higher frequencies we observe a dynamic phase transition.
Applying standard finite-size scaling techniques from the theory
of second-order equilibrium phase transitions to this
{\it nonequilibrium} transition,
we obtain estimates for the transition
frequency and the critical exponents
($\beta / \nu \approx 0.11$,
$\gamma / \nu \approx 1.84$ and $\nu \approx 1.1$).
In addition to their significance for the interpretation of recent
experiments on switching in ferromagnetic and ferroelectric nanoparticles
and thin films, our results provide
evidence for the relevance of universality and finite-size scaling to
dynamic phase transitions in spatially extended nonstationary systems.
\end{abstract}



\section{Introduction}
\label{sec_intro}

The term hysteresis comes from the Greek {\it husterein} 
($\stackrel{\scriptscriptstyle c}{\upsilon} \! \! \sigma \tau 
\epsilon \rho \acute{\epsilon} \omega$) 
which means ``to be behind'' \cite{EWIN1881}. 
It describes the lagging of
an effect behind its cause, as when the magnetization of
a body lags behind periodic changes in the applied field.
While the magnetization response of a ferromagnet
in an oscillating field is probably the example most familiar
to physicists and engineers 
\cite{WARB1881,EWIN1882,stei1892,KLIN92,maye91,ahai96},
hysteresis is a quite common phenomenon.
For instance, it is also seen in ferroelectrics 
\cite{ishi71,orih92,DUIK90,beal94,rao91,mito94,orih94,hash94},
in which
the polarization lags behind a time-varying electric field.
Other examples of hysteresis include electrochemical adsorbate
layers that are driven through a phase transition by an oscillating 
electrode potential in a Cyclic Voltammetry experiment
\cite{bard80,rikv95} and liquid-crystalline
systems driven through a phase transition
by pressure oscillations \cite{chen96}.
Recently, a new class of superconducting materials
including Dy${\rm Ni}_{2}$${\rm B}_{2}$C \cite{peng98} have shown
hysteresis in the resistivity when
subject to an oscillating magnetic field.
Acoustic hysteresis in crystals \cite{burl97} occurs when the
ultrasonic absorption coefficient changes due to oscillations in the
amplitude of an ultrasonic wave.

In recent years 
new experimental techniques, such as magnetic force microscopy (MFM)
\cite{mart87,chan93,lede93,lede94,lede94-2}, have been developed that 
permit measurements of the magnetization state and switching behavior 
of particles as small as a few nanometers. Ferromagnetic particles in 
this size range consist of a single domain in equilibrium.
Together with ultrathin films, they are of interest as potential
materials for ultra-high density recording media. 
The dynamics of magnetization reversal in nanoscale systems has been
modeled by kinetic Ising systems subject to sudden field reversal 
\cite{rich95,rich96,rich96-2,kole97,kole97-MRS,geilo98}.
These numerical and analytical studies give results in qualitative agreement 
with the experiments mentioned above.
Recent experiments on ultrathin ferromagnetic Fe/Au (001)
films \cite{he93} and thin p(1$\times$1) Fe films on W(110) \cite{suen97}
have considered the frequency dependence of
hysteresis loop areas, which were interpreted in terms of effective
exponents consistent with those found for a continuous spin model 
\cite{rao91,rao89,rao90,rao90_2}. 
Similar experiments have been performed on ultrathin Co films on Cu(001). 
A study of this system by Jiang {\it et al.\/} \cite{jiang95} 
reported exponents consistent with a mean-field treatment of the Ising model, 
whereas a recent study by Suen {\it et al.\/} \cite{SUEN99} finds very 
small effective exponents in the low-frequency regime, 
apparently consistent with the theoretical results we report here. 

The above discussion is far from
an exhaustive account of hysteresis examples,
but it does give an idea of the diversity of situations
in which this nonlinear, nonequilibrium phenomenon is important.
Systems that exhibit hysteresis have in common a nonlinear, 
irreversible response which lags behind the applied force.
Numerous general mathematical theories have been formulated to model
hysteretic behavior in a variety of systems, including
the Preisach model \cite{maye91,visi94,brok96,prei35} and
systems of differential equations that display
discontinuous bifurcations \cite{visi94,brok96,nayf95}.
Hysteresis in thermodynamic systems is often
due to the presence of a first-order phase transition, which is
the source of strong nonlinearity in the system.
These details of the nonlinear response
can, however, be quite different in 
different systems and even in different parameter regimes for the same 
system.
The details must be carefully considered in order to
accurately predict such aspects of the hysteretic response as its dependence 
on the frequency, amplitude, and waveform of the oscillating force. 
Here we present a study of hysteresis in a particular 
model system which incorporates both spatial degrees of freedom and thermal 
fluctuations and which has a first-order equilibrium phase transition.
The system response in the parameter regime studied in this paper
is {\it self-averaging} and may be described by
the Kolmogorov-Johnson-Mehl-Avrami (KJMA) theory of metastable decay
\cite{kolm37,john39,avra39}.

Specifically, we consider hysteresis in a spin-$1/2$,
nearest-neighbor, kinetic Ising ferromagnet
on a two-dimensional square lattice with periodic boundary conditions,
which is subject to a sinusoidally oscillating field.
For convenience, and because many of the experimental measurements
of hysteresis involve magnetic systems, we use
the customary magnetic language in which the order parameter
is the dimensionless 
magnetization per site, $m(t) \in [-1,+1]$, and the force is the 
magnetic field $H(t)$. However, we expect our results also to apply 
to hysteresis in other areas of science. 
For example, in dielectrics $m(t)$ and $H(t)$ can be 
re-interpreted as polarization and electric field, in adsorption problems 
as coverage $\theta(t) = \left[ m(t) + 1 \right]/2$ \cite{error_md}
and (electro)chemical potential or (osmotic) pressure, etc.

Below its critical temperature
and in zero field, this model has two degenerate
ordered phases corresponding to a majority of the spins in the
positive or the negative direction.
A weak applied field breaks the degeneracy,
and the phase with the spins aligned (anti-aligned) with the field is 
stable (metastable). If the field varies periodically in time, the system 
is driven back and forth across a first-order phase transition
at $H \! = \! 0$, and the two
phases alternate between being momentarily stable and metastable.
As a result, $m(t)$ lags behind $H(t)$, and hysteresis occurs.
In the regime of large system size,
moderately strong field, and temperature
well below $T_{c}$ considered here, the system switches smoothly and almost
deterministically between the two magnetized phases.

The metastable phase in Ising
models subject to a sudden field reversal from $H$ to $-H$
decays by different
mechanisms, depending on the magnitude of $H$, 
the system size $L$, and the temperature $T$.
Two distinct regimes are separated by 
a crossover field called the dynamic spinodal,
$H_{\rm DSP} \sim (\ln L)^{-1/(d-1)}$,
where $d$ is the spatial dimensionality \cite{tomi92A,rikv94}.
Detailed discussions of these different decay modes are found
in Refs.~\cite{geilo98,rikv94,rikv94_review}.
At sufficiently low $T$ that the single-phase correlation lengths are 
microscopic, the different decay regimes 
can be distinguished by the interplay
among four length scales: the lattice spacing $a$,
the system size $L$, the radius of a
critical droplet $R_{c}\propto 1/|H|$, and the average distance
a supercritical droplet interface propagates before encountering
another droplet $R_{0} \propto \exp \left \{ \Xi_{0}(T)/
\left [(d+1)|H|^{d-1} \right ] \right \}$.
The physical significance of $\Xi_{0}(T)$ is explained
in Sec.~\ref{sec_Model}.
In this paper, we specifically consider decay in the
multi-droplet (MD) regime
[$H_{\rm DSP}(T,L) < |H| < H_{\rm MFSP}(T)$ where 
$R_c(H_{\rm MFSP}) \approx 0.5$]. 
In terms of the characteristic lengths, the MD regime is defined by 
 \begin{equation}
 \label{eq_lengths}
  a \ll R_c \ll R_0 \ll L \;.
 \end{equation}
Here, the decay of the metastable phase proceeds by random
homogeneous nucleation of {\it many}
critical droplets of the stable phase, which then grow and coalesce.
The MD regime is distinct from the strong-field regime, 
$|H| > H_{\rm MFSP}(T)$. 
It is also distinct from the single-droplet (SD) regime
($|H| < H_{\rm DSP}(T,L)$), where $a \ll R_c \ll L \ll R_0$.
In the SD regime, the decay of the metastable phase proceeds by random
homogeneous nucleation of a {\it single}
critical droplet of the stable phase.
Hysteresis in that regime is described in detail
in Ref.~\cite{side98-SD}.
The present study, as well as our previous work
\cite{side9697,side98-MMM}, 
shows that the response to an oscillating field is
significantly different in the MD and SD regimes.

Theoretical and computational studies of hysteresis
have been performed for several models, using a variety of methods 
\cite{CHAK99}.
These include various
studies of models with a single degree of freedom, equivalent to
mean-field treatments of the Ising
model \cite{jung90,tome90,luse94}, Monte Carlo (MC)
simulations of the spin-1/2 Ising model
\cite{rao90,rao90_2,acha94_review,acha92-1,acha92-2,acha94,%
acha94-2,lo90,fan95-2,acha97,acha97-2,acha95},
and several $O(N)$ type models 
\cite{rao91,rao89,rao90,rao90_2,dhar92}.
These studies were performed
with variations in the details of the simulations
and in the model parameters.
Most of them indicate that the average hysteresis-loop area
appears to display
power-law dependences on the frequency and amplitude of $H(t)$. 
However, there is no universal agreement on the values of the exponents,
either experimentally or theoretically.
For the Ising model it has been pointed out that nucleation effects 
would lead to an asymptotically logarithmic frequency dependence 
\cite{beal94,kole97,rao90,thom93}.
A mean-field model exhibits a dynamic phase
transition in which the mean period-averaged magnetization
changes from a nonzero to zero mean value \cite{tome90}.
Such a dynamic phase transition has been observed in
MC simulations of a kinetic Ising model as well
\cite{acha94_review,acha92-1,acha92-2,acha94,acha94-2,lo90,acha97,%
acha97-2,acha95,side98-PRL}.
A fundamentally different example of criticality in a hysteretic system is
the zero-temperature, random-field Ising model, which exhibits critical
behavior in the hysteresis loop as a function of disorder \cite{seth93}.

The work presented in this paper and in 
Refs.~\cite{side98-SD,side9697,side98-MMM,side98-PRL} 
differs from most
past theoretical and numerical studies of hysteresis in two important ways.
First, mean-field models do not take into account
thermal noise and spatial variations
in the order parameter, thus ignoring fluctuations which may be important
in real materials. Second, most previous 
investigations of hysteresis in Ising models
have considered the frequency and amplitude dependence
of quantities such as
the loop area and the period-averaged magnetization
{\em without} considering the manner in which the metastable phase decays.
In this paper,
the long-time behavior of the hysteretic response is analyzed by
studying the power spectral densities of the magnetization time series
as well as the
statistical properties of the period-averaged magnetization $Q$,
the loop area $A$, and the correlation $B$.
These quantities are defined as follows.
 \begin{eqnarray}
 \label{eq_Q_md}
   Q & = &\frac{\omega}{2 \pi} \oint m(t) \ dt \\
 \label{eq_A_md}
   A & = & -\oint m(H) \ dH \\
 \label{eq_B_md}
   B & = & \frac{\omega}{2 \pi} \oint m(t) \ H(t) \ dt \;,
 \end{eqnarray}
where the initial time $t_0$ of the period is defined such that 
$H(t_0) = H(t_0 + 2 \pi / \omega) = 0$. 

Due to the multi-droplet decay mechanism, the average
hysteresis-loop area exhibits an extremely slow crossover to a logarithmic
decay with frequency and amplitude in the asymptotic low-frequency 
limit \cite{side98-MMM}.
This asymptotic behavior of the loop area in the MD regime is
qualitatively similar
to that for the SD regime for two dimensions \cite{side98-SD,side98-MMM}. 
However, the calculation is somewhat more involved, 
and the quantitative behavior
of the loop area is different for the two regimes. 

Beside our results on the low-frequency loop areas, our
most significant finding is
detailed evidence of finite-size scaling at
a dynamic phase transition (DPT) in the MD regime.
Here we provide a full account of these results, which we
briefly reported in Ref.~\cite{side98-PRL}.
This transition can be intuitively understood as a competition
between two time scales: the period of the external field,
$2 \pi / \omega$, and the average lifetime of the metastable phase,
$\left < \tau(H) \right >$,
defined as the first-passage time to a magnetization
of zero following an instantaneous field reversal from $H$ to $-H$.
If $2 \pi / \omega \ll \left < \tau(H_{0}) \right >$
[$H_{0}$ is the amplitude of $H(t)$]
the magnetization cannot fully switch sign
within a single period, and $|Q| > 0$.
We shall refer to this situation as the ordered dynamic phase.
If $2 \pi / \omega \gg \left < \tau(H_{0}) \right >$
the magnetization follows the field, and $Q \approx 0$.
This is the disordered dynamic phase.
Between these limits there is a critical frequency
at which $\left <|Q| \right >$ appears to
become singular in the infinite-system limit.
We emphasize that the DPT is a nonequilibrium phase transition and that
the probability distribution of the system magnetization
which characterizes the two phases never relaxes into a stationary state.
However, the ``filtered'' time series of $Q$ for successive field
periods is a stationary stochastic process.
To avoid confusion we establish the following terminology.
By the term ``dynamic phase'' we mean one of the qualitatively different
system responses separated by the DPT.
In contrast, the term ``phase'' by itself refers in the conventional
way to a uniform thermodynamic phase. 

The rest of this paper is organized as follows.
Details of the model and a brief review of relevant aspects of the
Avrami theory of metastable decay are given
in Sec~\ref{sec_Model}.
Time-series data for the magnetization
and the period-averaged magnetization
$Q$ are discussed in Sec.~\ref{sec_timeseries_md}.
In Sec.~\ref{sec_psd_md} we discuss the power spectral
densities obtained from the time-series data.
In Sec.~\ref{sec_avrami} we obtain an analytical result for the
time-dependent system magnetization during a single period of the
field, based on the droplet theory of nucleation and a time-dependent
extension of the Avrami theory.
Section \ref{sec_ab_md} contains an analysis of the hysteresis-loop area
$A$ and the correlation $B$.
This section includes MC data for the probability distributions and averages
of $A$ and $B$ along with
theoretical predictions based on the results in Sec.~\ref{sec_avrami}.
In Sec.~\ref{sec_q_md} we consider
the period-averaged magnetization $Q$, which
is the order parameter for the dynamic phase transition.
Details of the finite-size scaling analysis for this transition
are given in this section as well.
A summary, discussion, and topics for future study
are presented in Sec.~\ref{sec_md_conclusion}.

\section{Model}
\label{sec_Model}

The model used in this study is a kinetic, nearest-neighbor
Ising ferromagnet on a
hypercubic lattice with periodic boundary conditions.
The Hamiltonian is given by
 \begin{equation}
 \label{eq_Hamil}
  {\cal H } = -J \sum_{ {\em \langle ij \rangle}} {\em s_{i}s_{j}} 
               - H(t) \sum_{i} {\em s_{i}},
 \end{equation}
where $H(t) \! = \! -H_{0} \sin (\omega t)$,
$s_{i} \! = \! \pm 1$
is the state of the $i$th spin,
$\sum_{ {\em \langle ij \rangle} }$ runs over all
nearest-neighbor pairs, and $\sum_{i}$ runs over all
$N \! = \! L^{d}$ lattice sites.
The magnetization per site is
 \begin{equation}
 \label{eq_m(t)}
  {m(t)} = \frac{1}{L^{d}} \sum_{i=1}^{N} {\em s_{i}(t)} \;.
 \end{equation}

The dynamic used
is the Glauber \cite{glau63} single-spin-flip Monte
Carlo algorithm with updates at randomly chosen sites.
The time unit is one Monte Carlo step per spin (MCSS).
The system is put in contact with a heat bath at temperature
$T$, and each attempted spin flip 
from ${\em s_{i}}$ to ${\em -s_{i}}$ is accepted with
probability \cite{bind88}
 \begin{equation}
 \label{eq_Glauber}
 W(s_{i} \rightarrow -s_{i}) = 
 \frac{ \exp(- \beta \Delta E_{i})}{1 + \exp(- \beta \Delta E_{i})} \ .
 \end{equation}
Here $\Delta E_{i}$ is the change in the energy of the system
that would result
if the spin flip were accepted,
and $\beta \! = \! 1/k_{\rm B}T$ where $k_{\rm B}$ is Boltzmann's constant. 
It has been shown in the weak-coupling limit that the 
stochastic Glauber dynamic can be
derived from a quantum-mechanical Hamiltonian
in contact with a thermal heat bath modeled as a collection
of quasi-free Fermi fields in thermal equilibrium \cite{mart77}.

The average number of droplets of the stable phase
that are formed per unit time and volume is given by the
field and temperature dependent nucleation rate per unit volume,
 \begin{equation}
 \label{eq_I}
  I \left ( H(t),T \right ) \approx B(T) |H(t)|^K 
                   \exp \left [- \frac{\Xi_{0}(T)}{|H(t)|^{d-1}} \right ].
 \end{equation}
The notation follows that of Ref.~\cite{rich95}, where $B(T)$ is a 
non-universal temperature dependent prefactor, and $K$ and $\Xi_{0}(T)$
are known from field theory 
\cite{lang6769,gnw80},
MC simulations \cite{rikv94} and
numerical transfer-matrix calculations \cite{ccga93,ccga94};
their values are listed in Table \ref{table_constants_md}.
The quantity $\Xi_{0}(T)$ is the field-independent part of the
free-energy cost of a
critical droplet, divided by $k_{B}T$.
The field, $H(t)$, is the only quantity through which
$I \left ( H(t),T \right )$ depends on time in this adiabatic approximation.

Several other quantities, whose values do not depend on the frequency
of the field, are required as input for the theoretical
calculations in the following sections and
are listed in Table \ref{table_constants_md}.
They are determined through what we refer to as
``field-reversal simulations.''
In these simulations the system initially has all spins up,
i.e.~positive.
It is then subjected to a static field of magnitude $H_0$ 
with a sign opposite the system magnetization.
This instantaneous
field quench prepares the system in a metastable state, and the
decay of this metastable phase proceeds by the MD mechanism outlined in
the introduction.

In the MD parameter regime the average size of a critical droplet $R_{c}$,
and the average distance between droplets $R_{0}$, are both much smaller
than the system size.
Therefore, many droplets nucleate and grow to
drive the system into the stable
phase, resulting in an almost deterministic decay process.
This can be understood by imagining the system subdivided into
cells of linear size $R_{0}$, which each contain a single droplet.
Each of these subsystems will appear to be in the SD regime, and the
time taken to nucleate a critical droplet is stochastic with an
exponential probability distribution.
However, as a consequence of the Central Limit Theorem
the probability density of the
lifetime for the entire system asymptotically
approaches a Gaussian as the system size increases.
The decay process thus becomes increasingly deterministic,
with a lifetime distribution
whose variance decreases as $L^{-d}$ with increasing system size \cite{rich95}.

In addition to the ``self-averaging'' process described above,
two other concepts are needed to understand the MD decay process:
the droplet interface velocity and the overlapping
of growing supercritical droplets.
Our detailed
treatment of these effects is given in Sec.~\ref{sec_avrami},
where the theory
of MD decay in static field is generalized to time-dependent fields.
Accounting for all of these effects, an expression may be obtained for
the time-dependent magnetization of a system in a field-reversal experiment.
In the KJMA approximation \cite{kolm37,john39,avra39},
it is assumed that the positions and sizes
of the growing droplets are uncorrelated.
In the simple field-reversal case this leads to the
well known ``Avrami's Law,''
 \begin{mathletters}
 \begin{eqnarray}
 \label{eq_avrami1_a}
  m(t) & = & 2 \exp \left [- \Phi(t) \right ] -1 \\
 \label{eq_avrami1_b}
       & = & 2 \exp \left [- \int_{0}^{t}
                     I \Omega_{d}(v_{0}t')^{d} dt' \right ] -1 \\
 \label{eq_avrami1_c}
       & = & 2 \exp \left [- \frac{\Omega_{d} v_{0}^{d} I}
                    {d+1} t^{d+1} \right ] - 1 \ ,
 \end{eqnarray}
 \end{mathletters}
where $v_{0}$, the
interface velocity for a growing droplet of the stable
phase, is assumed to be constant,
$\Omega_{d}$ is a proportionality constant such that the volume $V$
of a droplet of radius $R$ is $V \! = \! \Omega_{d} R^{d}$,
and the other constants have been introduced previously.
The integral $\Phi(t)$ in Eq.~(\ref{eq_avrami1_b}) is the
``extended volume'' \cite{avra39}, i.e.~the total volume fraction
of droplets of the equilibrium phase at time $t$,
{\it uncorrected} for overlaps. 
The assumption that the positions and sizes of the droplets are uncorrelated
leads directly to the exponential relation for $m(t)$ \cite{rikv85A}.
Solving Eq.~(\ref{eq_avrami1_c}) for the time at
which $m \! = \! 0$ gives the average lifetime in the MD regime,
 \begin{equation}
 \label{eq_tau_md}
  \left < \tau \right > =
    \left [ \frac{\Omega_{d} v_{0}^{d} I}
            {\ln 2 (d+1)} \right ]^{-\frac{1}{d+1}} \ ,
 \end{equation}
which, in contrast to the SD regime, is independent of $L$. 
To describe the hysteretic response in the MD regime
in Secs.~\ref{sec_avrami} and \ref{sec_ab_md} we employ
a time-dependent extension of this theory, in which $I$ and $v_{0}$
are both functions of time through $H(t)$.

\section{Time-series data}
\label{sec_timeseries_md}

\begin{figure}
\centerline{\psfig{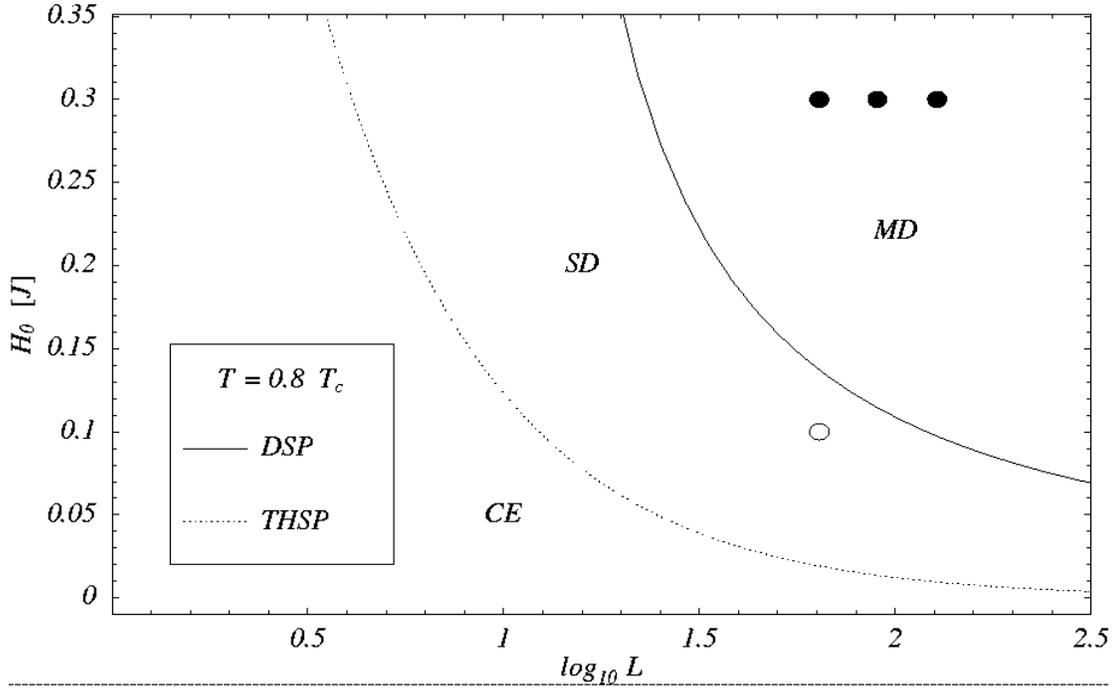}}
\caption[Location of MD simulations]
{\label{fig_hsw_L_md}
Location of MD simulations in the $(H_{0},L)$ plane.
The solid curve represents the dynamic spinodal (DSP),
$H_{\rm DSP} \sim (\ln L)^{-1/(d-1)}$.
This theoretical curve is an asymptotic result obtained by
setting $R_{0} \approx L$.
The filled circles denote the system sizes
($L \! = \! 64$, $90$, $128$)
and field amplitude ($H_{0} \! = \! 0.3J$) used here to study
hysteresis in the MD regime.
The open circle denotes the system size and field amplitude used to study
hysteresis in the SD regime in Ref.~\protect\cite{side98-SD}.
The dotted curve represents a theoretical result
for the ``thermodynamic spinodal field'' (THSP) \protect\cite{tomi92A,rikv94}
which separates
the SD region from the coexistence regime (CE).
It is obtained by setting $R_{c} \approx L$.
}
\end{figure}

All numerical simulations reported in this paper
are performed for $d \! = \! 2$, $T \! = \! 0.8 T_{c}$
and one of three system sizes, $L \! = \! 64$, $90$, and 128.
A sinusoidal field is applied to the system
with amplitude, $H_{0} = 0.3J > H_{\rm DSP}(L)$, chosen such that in
field-reversal simulations the system is clearly in the MD regime for a
field of magnitude $H_{0}$ for all three values of $L$.
This is illustrated in Fig.~\ref{fig_hsw_L_md}.
The dynamic spinodal field is approximated by $H_{\rm DSP} \approx H_{1/2}$,
where $H_{1/2}$ is the value of $H$ (for given $L$ and $T$) for which
the relative standard deviation of the lifetime,
$r \! = \! \sigma_{\tau}/\left< \tau \right>$ is $1/2$
($\sigma_{\tau}$ is the standard deviation of the lifetime).
The values of $H_{\rm DSP}$ and
$\left < \tau \right >$
for $T \! = \! 0.8T_{c}$ and $L \! = \! 64$ \cite{ramo97,jlee95}
are given in Table \ref{table_constants_md}.
For the larger systems,
$\left < \tau \right >$ is approximately unchanged while $H_{\rm DSP}$
and $r$ are smaller than for $L \! = \! 64$.
Thus the system is well within the MD regime
for all three sizes used.

To obtain the raw time-series data, the system was initially prepared
with either a random arrangement of up and down spins with $m(t=0) \approx 0$,
or with a uniform arrangement with all spins up.
Then the sinusoidal field was applied and
changed every attempted spin flip,
allowing for a smooth variation of the field.
The time series did not appear to depend on the initial conditions after
a few periods.
For each system size, the simulations were performed with several values of
the driving frequency $\omega$.
For each frequency, we recorded the time-dependent magnetization $m(t)$.
Most of the simulations at intermediate and high frequencies were
recorded for
approximately $16.9 \times 10^{6}$ MCSS.
(Simulations for some of the lowest frequencies were recorded for
approximately $5.9 \times 10^{5}$ MCSS.)
Files containing the data for these longest runs are about $800$ megabytes
and required $9$ days (one month) to run for $L \! = \! 64$ 
($L \! = \! 128$) on a
single $66$~MHz node of an IBM sp2 computer.
Since the hysteresis depends on the competition between the two time scales
represented by the field period and the metastable lifetime,
we chose the frequencies of $H(t)$ by specifying the ratio
 \begin{equation}
 \label{eq_R}
   R = \frac{(2 \pi / \omega)}{\left < \tau(H_{0}) \right >} \ .
 \end{equation}
One may think of $R$ as a scaled period and $1/R$ as a scaled frequency.

We note that $\left < \tau(H_{0}) \right >$ is the
``shortest of the long time scales'' that describe the system.
For $T \! = \! 0.8T_{c}$ and all the values of $L$ used here,
the time scale for spontaneous fluctuations between the phases in the
absence of an applied field, $\left < \tau(0) \right >$, is
essentially infinite.
Even when the field has its maximum strength $H_{0}$,
the nucleation of the critical droplets necessary to
leave the metastable phase is always driven by the thermal fluctuations.
Driving the system from the metastable to the stable phase therefore
truly depends on the joint action of the random thermal noise and the
deterministic oscillating field.
Figure \ref{fig_mt_md} shows short initial segments of the magnetization
time series for four different values of $R$.
The figures for the first three values of $R$ are chosen to represent:
a system in the ordered dynamic
phase ($R \! = \! 3$, $\left < |Q| \right > > 0$),
near the dynamic transition ($R \! = \! 3.436 \approx R_{\rm cr}$), and
in the disordered dynamic
phase ($R \! = \! 7$, $\left < |Q| \right > \approx 0$).
This value for $R_{\rm cr}$ is obtained by finite-size scaling
analysis of the probability density for $Q$, as described in
Sec.~\ref{sec_q_md}.
The time-series segment shown for $R \! = \! 200$ is deep in the disordered
dynamic phase region
and illustrates the behavior of the system for very low frequencies.
The standard deviation of the average lifetime in the MD regime is relatively
small compared to the SD regime.
If the period of $H(t)$ is sufficiently long,
the system has enough time to switch phases during a single half-period.
This is clearly seen at $R \! = \! 7$, for which the system switches
during practically every half period.
Soon after $m(t)$ reaches saturation, the field passes through zero and
favors the opposite phase.
Similar behavior is seen in the time series for $R \! = \! 200$, except that
the period is so long that the system decays to the favored phase
well before the field reaches its maximum value.
Then the magnetization fluctuates near its equilibrium value 
$m(t) \approx \pm 1$ until the
field again switches sign and the system once more becomes metastable.
If the period of $H(t)$ is sufficiently short,
the system does not have time to switch during a single half-period.
This can be seen for $R \! = \! 3$.
While the field favors the opposite phase, the magnetization changes as
many critical droplets nucleate and begin to grow.
Before the magnetization
can completely reverse however, the field changes
sign and the droplets of the now unfavored phase shrink and disappear.
For $R \! = \! 3.436$, the period near the critical value, the 
period-averaged magnetization slowly ``meanders'' from positive to 
negative values over several periods of $H(t)$.
This ``slow switching'' occurs many times over the entire time series.
The number of field cycles shown in Fig.~\ref{fig_mt_md} is small compared
to the total number of cycles in an entire time series.

The ``slow switching'' seen near the
dynamic phase transition also occurs
for frequencies in the ordered
dynamic phase region, where
$\left < |Q| \right > > 0$.
However, there the times between consecutive switches
are too long to show in plots of $m(t)$ vs $t$.
For this reason, the ``filtered time series'' for $Q$
are shown in Fig.~\ref{fig_qt_md}, which provide plots of $Q$
for consecutive periods in the magnetization time series.
Even for the $Q$ time series, the number of periods shown is small compared
to the total number of periods, except for
$R \! = \! 200$ which displays the entire time series.
For the low frequencies, the $Q$ values are concentrated near $Q \! = \! 0$,
with larger fluctuations for $R \! = \! 7$ than for $R \! = \! 200$.
Analysis of the $Q$ data and the dynamic phase transition will be
detailed in Sec.~\ref{sec_q_md}.

\section{Power spectral densities}
\label{sec_psd_md}

A standard method used to characterize a time
series is to calculate its power spectral density (PSD).
Figure \ref{fig_psdMD} shows the PSDs of the raw data,
short segments of which are shown in Fig.~\ref{fig_mt_md}.
The technical details on how the PSDs were obtained are elaborated
in Ref.~\cite{side98-SD}. For clarity, 
the PSDs for different driving frequencies,
shown in Fig.~\ref{fig_psdMD}(a) with $L \! = \! 64$,
have been shifted in the vertical direction by arbitrary offsets.
The same spectra are plotted in Fig.~\ref{fig_psdMD}(b) with no offset.
The fourth spectrum shown in Fig.~\ref{fig_psdMD}, labeled ``background,''
corresponds to thermal equilibrium fluctuations in a single 
thermodynamic phase.
To obtain this spectrum, a simulation was performed
on a system with the same size, temperature, and for
the same number of MCSS as the other spectra, in
a {\it static} field of $H_{0}/\sqrt{2}$.

To describe the PSD for each frequency, we identify three distinct regions:
1) the peaks,
2) the thermal noise region, and
3) the low-frequency region.
The most prominent features of the PSDs are the sharp peaks.
For $R \! = \! 3$ and $3.436 \approx R_{\rm cr}$,
the first peak in the spectrum corresponds to
$\omega$, the frequency of the external field.
The second peak corresponds to $2 \omega$, and so on.
These odd and even harmonic peaks arise because the shape of the
time series is not purely sinusoidal due to the nonlinear response of the
system.
For $R \! = \! 7$ (and longer periods),
only odd harmonics are seen.
The extinction of the even harmonic peaks occurs
because the shape of the time
series is beginning to resemble a square wave (see Fig.~\ref{fig_mt_md}).
The power $p_{n}$ contained in the $n$th component of the Fourier series
for a pure square wave is
$p_{n} = 16 [\sin (n \pi/2)]^4/(n \pi)^{2}$, which decays as $n^{-2}$ 
and vanishes for even $n$. This enables one to understand the
reduced second harmonic in the PSD for $R \! = \! 7$,
which is just barely observable between the first two sharp peaks.
However, in contrast to our observations in the SD regime \cite{side98-SD},
no dips in the PSD at even $n$, corresponding to the zeros of $p_{n}$
\cite{shne94-1-3}, 
were observed for the values of $R$ analyzed here.

Unlike the SD regime \cite{side98-SD},
the highest frequencies for each of the PSDs 
do not fall onto the thermal noise background.
Since the average lifetime in the MD regime is much smaller than in the 
SD regime,
the $R$ values shown in Fig.~\ref{fig_psdMD} correspond to much larger
frequencies in units of ${\rm MCSS}^{-1}$.
Therefore, the time scales characterizing the peaks and the thermal 
noise regions
are not as well separated as for the SD regime,
so the two regions overlap.
Also, none of the PSDs shown here are for low enough frequencies that
the metastable phase can decay and the system remain
in the stable state sufficiently long during each half-period
to sample the purely thermal fluctuations which would display
exponential time correlations.

The low-frequency region comprises the portion of each spectrum between the
first peak and the lowest resolved frequency.
The PSD in this region exhibits a strong dependence on the
frequency of the field.
Significant amounts of power in this portion of a PSD indicate the presence
of slow behavior on time scales larger than that of the driving field.
Near the transition frequency, $R = 3.436 \approx R_{\rm cr}$,
the overall slope in the low-frequency region is close to $-2$.
This suggests that the long-time correlations are exponential
with a very large correlation time \cite{acha98}.
The turnover in the corresponding
Lorentzian PSD is not observable because of the lack of
low-frequency resolution due to the finite length of the time series.
For $R \! = \! 3$, the low-frequency region of the PSD also
suggests a Lorentzian, again with a correlation time that is difficult
to estimate due to the poor low-frequency resolution.
For $R \! = \! 7$, 
the flat low-frequency region is that of the PSD of white noise.
This is consistent with the behavior of the 
$Q$ time series shown in Fig.~\ref{fig_qt_md}(d).
The PSDs for other system sizes display a qualitatively similar
frequency dependence in the sharp peak and low-frequency regimes
discussed above.
However, there is a systematic size dependence in the
PSDs for the thermal noise background, which is smaller for larger $L$. 
This is easily understood since the variance of the magnetization in
equilibrium should be proportional to $L^{-2}$.

\section{Derivation of $m(t)$ from Avrami's Law}
\label{sec_avrami}

The theoretical predictions for the frequency dependence of
both the hysteresis-loop areas and the correlation rely on
numerically solving an expression for $m(t)$ during a single field
period.
This expression is derived from a generalization
of Avrami's law with time-dependent nucleation rate and growth velocity. 
Using these values for $m(t)$ in Sec.~\ref{sec_ab_md}
we explicitly calculate the average hysteresis-loop area,
$\left < A \right > \! = \!  -\left<\oint m(H) \ dH \right > $,
and the average correlation,
$\left < B \right >  \! 
= \! (\omega/2 \pi) \left < \oint m(t) \ H(t) \ dt \right >$.
Avrami's law gives the volume fraction of the metastable state
for systems in
which many non-interacting droplets nucleate, then grow and
coalesce without changing their shapes as the system decays to the 
stable phase.
The volume fraction of metastable phase is related to the magnetization
by $\phi(t) \approx \left [m(t)+1 \right]/2$.
The original KJMA calculation
shows that the volume fraction of the metastable phase, $\phi(t)$, is given by
Eq.~(\ref{eq_avrami1_a}).
For a homogeneous, time-dependent nucleation rate,
the extended volume in Eq.~(\ref{eq_avrami1_b}) is generalized to
 \begin{equation}
 \label{eq_Phi(t)_1}
  \Phi(t) = \int^{t}_{0} I(L,T;t_{\rm n}) V(t,t_{\rm n}) dt_{\rm n} \ ,
 \end{equation}
where $V(t,t_{\rm n})$ is the volume at time $t$
of a droplet which was
nucleated at time $t_{\rm n}$, and $I(L,T;t_{\rm n})$ is the time-dependent
nucleation rate given by Eq.~(\ref{eq_I}).
The volume $V(t,t_{\rm n})$ is given by
 \begin{equation}
 \label{eq_V(t)1}
  V(t,t_{\rm n}) = \Omega_{d} \left [ \int^{t}_{t_{\rm n}} v(t') dt'
                    \right ]^{d} \ .
 \end{equation}
The Lifshitz-Allen-Cahn approximation \cite{gunt83,lifs62,alle79}
is used to specify the interface velocity of a growing droplet as
$v(t) \approx \nu |H(t)|$.
(The proportionality factor $\nu$ should not be confused with the
critical exponent $\nu$, discussed in Sec.~\ref{sec_q_md}.)
The effects of the dependence of the proportionality factor
$\nu$ on the droplet radius,
which are remarkably minor, are discussed by Shneidman
and collaborators \cite{shne93,shne98}.
Equation (\ref{eq_Phi(t)_1}) with
$d \! = \! 2$ and $K \! = \! 3$ in Eq.~(\ref{eq_I}) then gives
 \begin{eqnarray}
  \Phi(t)
   \label{eq_Phi(t)_2} 
    & = & \int^{t}_{0} I(L,T;t_{\rm n})
         \Omega_{2} \left [ \int^{t}_{t_{\rm n}}
         \nu H_{0} \sin \omega t' dt' \right ]^2
         dt_{\rm n} \\
   \nonumber
   & = & \int^{t}_{0} I(L,T;t_{\rm n})
         \frac{\Omega_{2} \nu^{2} H_{0}^{2}}{\omega^{2}}
         \left [ \cos \omega t_{\rm n} - \cos \omega t \right ]^2
         dt_{\rm n} \\
   \nonumber
   & = & \frac{B(T) \Omega_{2} \nu^{2} H_{0}^{2}}{\omega^{2}} 
         \int^{t}_{0}
         \left [ \cos \omega t_{\rm n} - \cos \omega t \right ]^2
         |H_{0} \sin \omega t_{\rm n}|^{3} \\ 
   \nonumber
   &   & \times 
         \exp \left [-\frac{\Xi_{0}(T)}{|H_{0} \sin \omega t_{\rm n}|} \right ]
         dt_{\rm n} \ .
 \end{eqnarray}
Numerically integrating Eq.~(\ref{eq_Phi(t)_2}) for
$t \in [0,\pi / \omega]$
and substituting into Eq.~(\ref{eq_avrami1_a}) enables one to find
$\phi(t) \! = \! \exp \left[ -\Phi(t) \right]$
for a system which starts with $\phi(0) \! = \! 1$.
The only quantity which must be determined from the
MC simulations is $B(T)$,
the field-independent part of the prefactor in the nucleation rate.
In the present paper, this single parameter is set by demanding
that the average simulated loop area for $R \! = \! 200$ matches
the theoretical prediction (see Sec.~\ref{sec_ab_md}).

The integration of Eq.~(\ref{eq_Phi(t)_2}) can only be performed for the 
first half-period
of the field cycle, in which $m(t)$ and $H(t)$ have opposite signs.
When $m(t)$ and $H(t)$ have the same sign, the droplets that were formed
during the previous half-period will shrink.
In that case, the magnetization is unable to switch sign
during a single period
for sufficiently high frequencies
(see Fig.~\ref{fig_MDdroplet_schematic}(a)).
The KJMA theory cannot be used to find the volume fraction of
{\it shrinking} droplets of the metastable phase.
Therefore, an approximate calculation of $m(t)$ for
$t \in [\pi / \omega, 2 \pi / \omega]$
must be devised.
Assume that the shrinking droplets in Fig.~\ref{fig_MDdroplet_schematic}(c)
are described by merely reversing the nucleation and growth
process represented in Fig.~\ref{fig_MDdroplet_schematic}(b).
Then, the volume fraction of the growing, now stable, background for
$t \in [\pi / \omega, 2 \pi / \omega]$ would be
given by $\phi \left[(2 \pi / \omega) - t \right ]$.
In addition to the growing stable background, which encroaches upon the
droplets of the metastable phase during the second half period,
we assume that droplets of the stable phase nucleate and grow within
the shrinking metastable droplets as depicted in 
Fig.~\ref{fig_MDdroplet_schematic}(c).
Furthermore, we assume that this nucleation and growth process is also
described by Avrami's law, with the shrinking metastable phase analogous
to the metastable background filling the entire system
for $t \in [0, \pi / \omega]$.
The complete prescription for the volume fraction of the 
growing stable background for $t \in [\pi / \omega, 2 \pi / \omega]$
is then
 \begin{eqnarray}
  \nonumber
  \phi'(t)
   & \approx & \phi \left( \frac{2 \pi}{\omega} - t \right ) 
   + \Bigg\{ [{\rm shrinking \ metastable \ phase}] \\ \nonumber
   &   & \times
           [{\rm growing \ stable \ phase \ inside \ \
          shrinking \ droplets}] \Bigg\} \\
   \nonumber
   & = & \phi \left( \frac{2 \pi}{\omega} - t \right ) + 
        \left [1- \phi \left( \frac{2 \pi}{\omega} - t \right ) \right ]
        \left [1-\phi\left (t- \frac{\pi}{\omega}\right ) \right ] \\
   \label{eq_phiPrime}
   & = & 1 - \phi\left(t- \frac{\pi}{\omega}\right)
         \left [1- \phi\left(\frac{2 \pi}{\omega} - t \right) \right ] \ .
 \end{eqnarray}
Thus the theoretical result for the magnetization during an entire period
of $H(t)$ is
 \begin{equation}
 \label{eq_m(t)2periods}
  m(t) \approx \left \{ \begin{array}{ll}
      2 \phi(t) - 1 \,,   & 0 < t < \pi / \omega \\
      2 \phi'(t)- 1 
                 = 1 - 2 \phi\left(t- \frac{\pi}{\omega}\right) \\
		 \times
                 \left [1- \phi\left(\frac{2 \pi}
		 {\omega} - t \right) \right ] \,,
                      & \pi / \omega < t < 2\pi / \omega \ .
      \end{array} \right.
  \end{equation}
Note that, in this approximation $m(2 \pi/\omega) \! = \! m(0)$ for all
frequencies.
Although it gives very good values for $\left < A \right >$
at all frequencies and for
$\left < B \right >$ especially at low frequencies,
it gives a continuous change in $\left < Q \right >$
with $\omega$, with no sign of a dynamic phase transition.

It is possible to rewrite Eq.~(\ref{eq_Phi(t)_2})
for $\Phi(t)$ by making the substitution
$H \! = \! H_{0} \sin(\omega t)$ and $H' \! = \! H_{0} \sin(\omega t_{\rm n})$.
After tedious, but straightforward algebra this gives
 \begin{eqnarray}
 \label{eq_Phi(H)}
  \nonumber
  \Phi(H) 
    & = & \frac{B(T) \Omega_{2} \nu^{2} H_{0}}{\omega^{3}}
          \int^{H}_{0} H'^{3} e^{-\Xi_{0}(T)/H'} \\
    &   & \times \Bigg[ \sqrt{1- \left (\frac{H'}{H_{0}} \right )^2} -
         2 \sqrt{1- \left (\frac{H }{H_{0}} \right )^2} \\ \nonumber
    &   & + \frac{ 1- \left (\frac{H }{H_{0}} \right )^2}
         {\sqrt{1-\left (\frac{H'}{H_{0}} \right )^2}}
         \Bigg] dH' \ .
 \end{eqnarray}
The equation is valid for $H \in [0,H_{0}]$, so
this expression allows a theoretical calculation of the magnetization
for sufficiently low frequencies
such that $m(t)$ completely switches sign within
a {\it quarter} period of the field.
Notice that $\omega$ is no longer part of the integration operation.
A similar calculation has been used to describe
experimental data for switching currents in (Nb,Co)-doped
BaTi${\rm O}_{3}$ ceramics \cite{mito94}
and hysteresis loops for triglycine sulphate (TGS)
single crystals \cite{orih94,hash94}
in the presence of an oscillating electric field.
Those studies assume heterogeneous nucleation and therefore obtain
a frequency dependence of $\omega^{-d}$, rather than $\omega^{-(d+1)}$
[see Eq.~(\ref{eq_Phi(H)})] 
for the case of homogeneous nucleation studied here.
The integral in Eq.~(\ref{eq_Phi(H)}) is evaluated {\it once} for
$H \in [0,H_{0}]$, from which $m(t)$ for $t \in [0,\pi / (2 \omega)]$ at any
frequency can be calculated.
For numerical reasons, Eq.~(\ref{eq_Phi(H)})
enables the theoretical prediction
of the hysteretic response to be more easily calculated
for lower frequencies
than by using Eq.~(\ref{eq_Phi(t)_2}) alone.
The values of $m(t)$ obtained from both
Eq.~(\ref{eq_Phi(t)_2}), for intermediate to high frequencies,
and Eq.~(\ref{eq_Phi(H)}), for low frequencies,
are used
in calculating the theoretical predictions for
$\left < A \right >$ and $\left < B \right >$
in Sec.~\ref{sec_ab_md}.

\section{Hysteresis-loop areas and correlation}
\label{sec_ab_md}

In this section we calculate
the hysteresis-loop area $A$ and the correlation between the
field and the system magnetization $B$,
defined in Eqs.~(\ref{eq_A_md}) and (\ref{eq_B_md}) respectively.
These quantities are calculated over each period in the entire time series.
From the resulting ``filtered'' time series we construct histograms to
obtain the probability densities of $A$ and $B$ for each separate frequency
of the field.
The hysteresis-loop area represents the energy dissipated during a
single period of the field \cite{WARB1881,EWIN1882,stei1892}.
It is therefore one of the most important physical quantities characterizing
hysteretic systems, and it is frequently measured in experiments.

Recent experiments on
ultrathin ferromagnetic films \cite{he93,suen97},
as well as numerical simulations of two-dimensional
Ising models \cite{rao90,lo90,acha95,seng92}, have been
interpreted in terms of a low-frequency power law,
$A \propto H_{0}^{a} \omega^{b}$, with
a range of exponent values reported \cite{lo90,acha95,seng92}.
This interpretation is not fully consistent with the
fluctuation-free mean-field result \cite{jung90,luse94},
$A = A_{0} + {\rm const} [\omega^{2} (H_{0}^{2} - H_{\rm sp}^{2})]^{1/3}$
with positive constants
$A_{0}$ and $H_{\rm sp}$, which has been applied to analyze
experiments on ultrathin films of Co on Cu(001) \cite{jiang95}.
Nor does the single power-law dependence
agree with the logarithmic dependence expected if
thermally activated nucleation \cite{beal94,kole97,rao90,thom93} is the
rate-determining process.
Here we present in detail analytical and numerical results
that indicate a resolution of this puzzling situation.
A brief summary
of some of the results was given in Ref.~\cite{side98-MMM}.

\begin{figure}
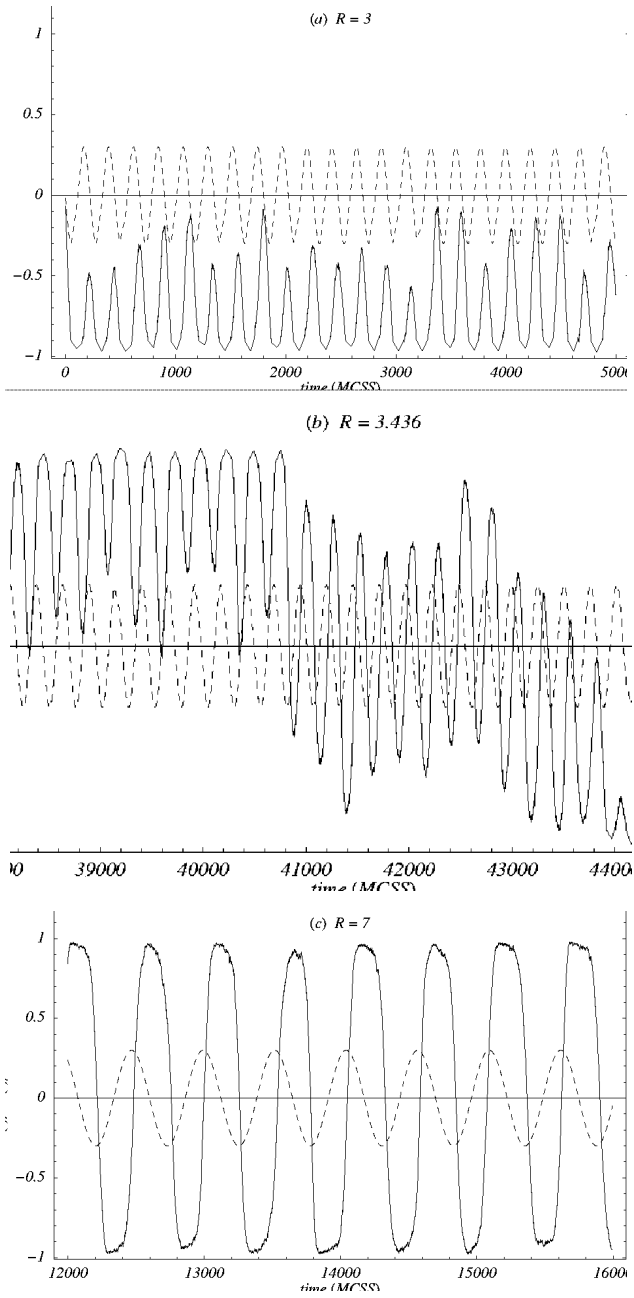
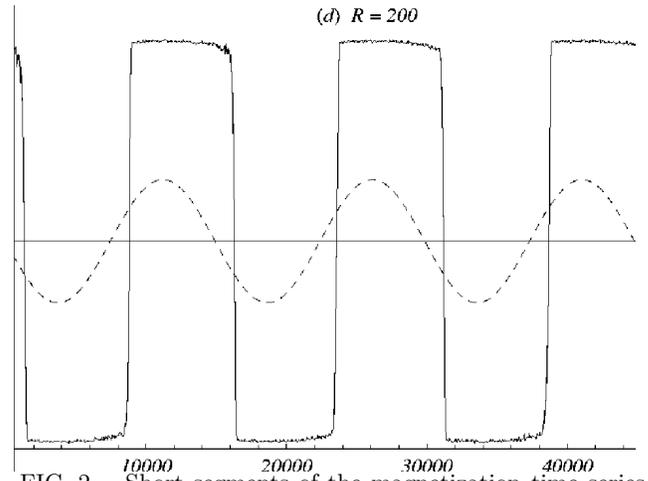

\centerline{\psfig{figure=mtMD_A_fig2a.epsi,width=3.25in,angle=270}}
\vspace{0.1in}
\centerline{
\psfig{figure=mtMD_B_fig2b.epsi,width=3.25in,angle=270}}
\vspace{0.1in}
\centerline{\psfig{figure=mtMD_C_fig2c.epsi,width=3.25in,angle=270}}
\vspace{0.1in}
\centerline{
\psfig{figure=mtMD_D_fig2d.epsi,width=3.25in,angle=270}}
\caption[MD time series]
{\label{fig_mt_md}
Short segments of the magnetization time series $m(t)$ (solid curve) and
the external field $H(t)$ (dashed curve) vs time $t$ for
$T \! = \! 0.8T_{c}$, $d \! = \! 2$, $L \! = \! 64$, and $H_{0} \! = \! 0.3J$.
The total length of the time series is approximately
$16.9 \times 10^{6} {\rm MCSS}$.
For these parameter values the average lifetime in static field
is $\left < \tau (H_{0}) \right > \approx 75$ MCSS.
The time series are shown for the scaled field periods
(a) $R \! = \! 3$,
(b) $R = 3.436 \approx R_{\rm cr}$,
(c) $R \! = \! 7$, and
(d) $R \! = \! 200$.
}
\end{figure}

\begin{figure}
\centerline{\psfig{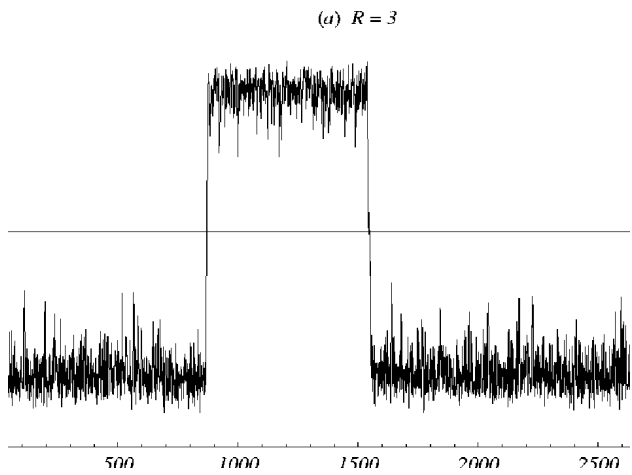}}
\vspace{0.1in}
\centerline{\psfig{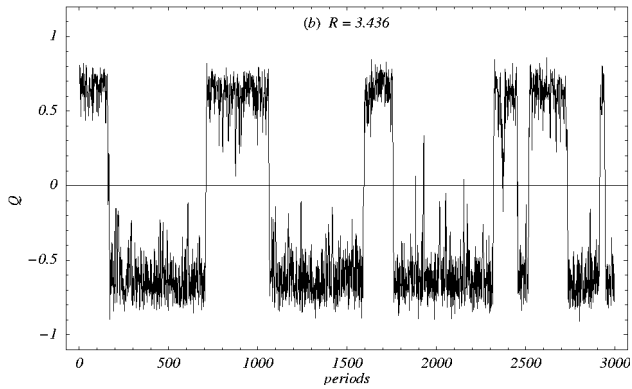}}
\vspace{0.1in}
\centerline{\psfig{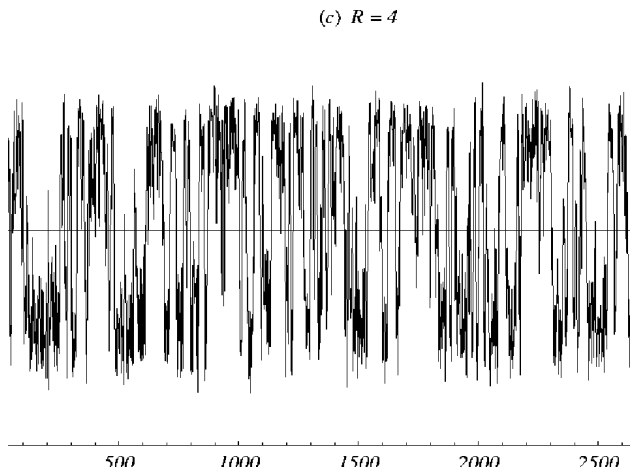}}
\vspace{0.1in}
\centerline{\psfig{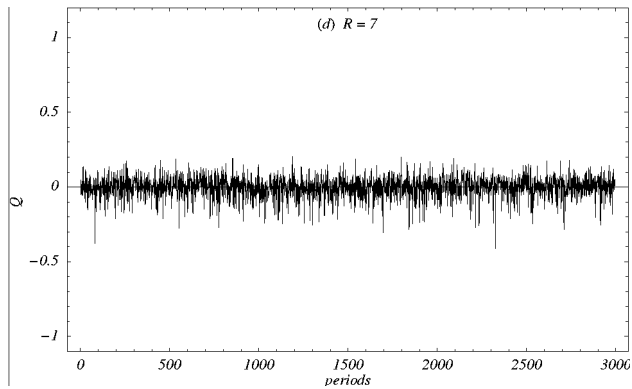}}
\vspace{0.1in}
\centerline{\psfig{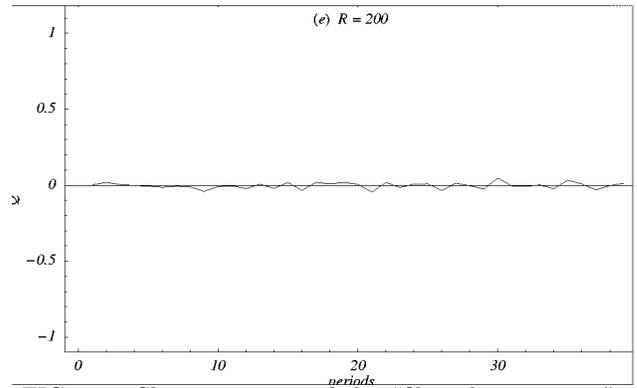}}
\caption[MD time series of period-averaged magnetization]
{\label{fig_qt_md}
Short segments of the ``filtered time series''
of period-averaged magnetization values,
$Q$, vs number of field periods.
The parameters used are the same as in Fig.~\protect\ref{fig_mt_md}.
(a) $R \! = \! 3$,
(b) $R = 3.436 \approx R_{\rm cr}$,
(c) $R \! = \! 4$,
(d) $R \! = \! 7$, and
(e) $R \! = \! 200$. 
}
\end{figure}

\begin{figure}
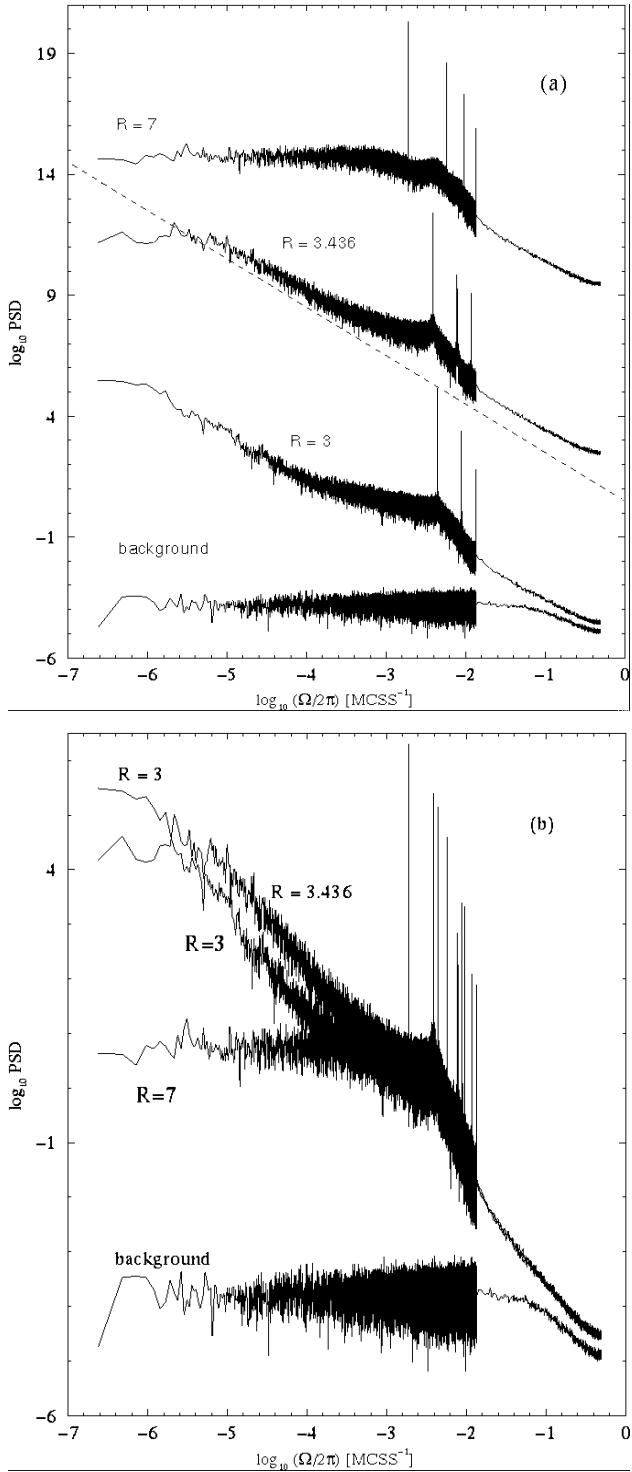

\centerline{\psfig{figure=psdMD_fig4a.epsi,width=3.25in,height=3.7in}}
\vspace{0.1in}
\centerline{\psfig{figure=noshift_L64_fig4b.epsi,width=3.25in}}
\caption[Power spectral densities (MD regime)]
{\label{fig_psdMD}
(a)
Power spectral densities (PSDs) for $L \! = \! 64$.
Spectra are shown for three different frequencies of the
field, and are plotted with an arbitrary offset for clarity.
In addition to a change in the amount of smoothing,
the right-hand section of each spectrum contains only one
out of every 250 data points to facilitate plotting.
The magnetization is sampled every 1.0 MCSS, so the Nyquist
frequency is $(\Omega_{\rm N}/2 \pi) \! = \! 0.5$ MCSS$^{-1}$.
The lowest frequency that can be resolved is
$2.38 \times 10^{-7} \ {\rm MCSS}^{-1}$.
The dashed line with slope $-2$ is a guide to the eye.
(b)
Same spectra without the offset
to illustrate how all three PSDs fall near the
thermal noise background at high frequencies.
Note that the spectra for $R \! = \! 3$ and $R = 3.436 \approx R_{\rm cr}$
cross at low frequencies.
}
\end{figure}

\begin{figure}
\centerline{\psfig{figure=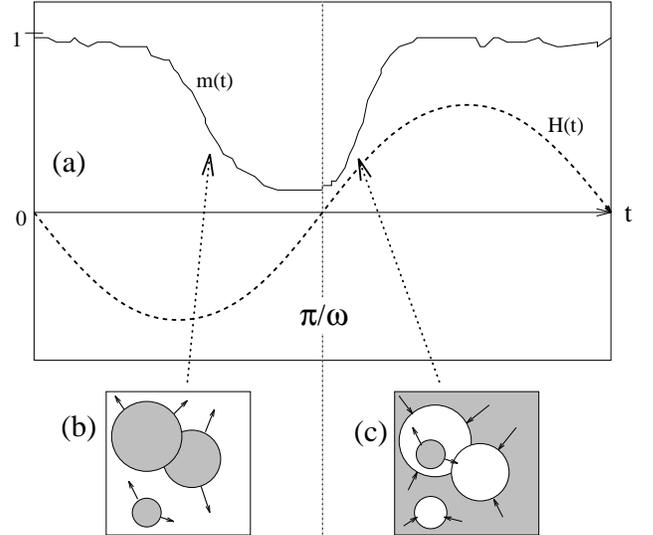,width=3.25in}}
\caption[MD droplet schematic]
{\label{fig_MDdroplet_schematic}
Schematic of growing and shrinking droplets in
the MD regime for a sufficiently high frequency such that
$m(t)$ does not completely switch during a period.
In (b) and (c) the dark regions represent the stable phase, 
and the light regions represent the metastable phase.
The arrows indicate the growth direction of the droplet interfaces.
}
\end{figure}

\begin{figure}
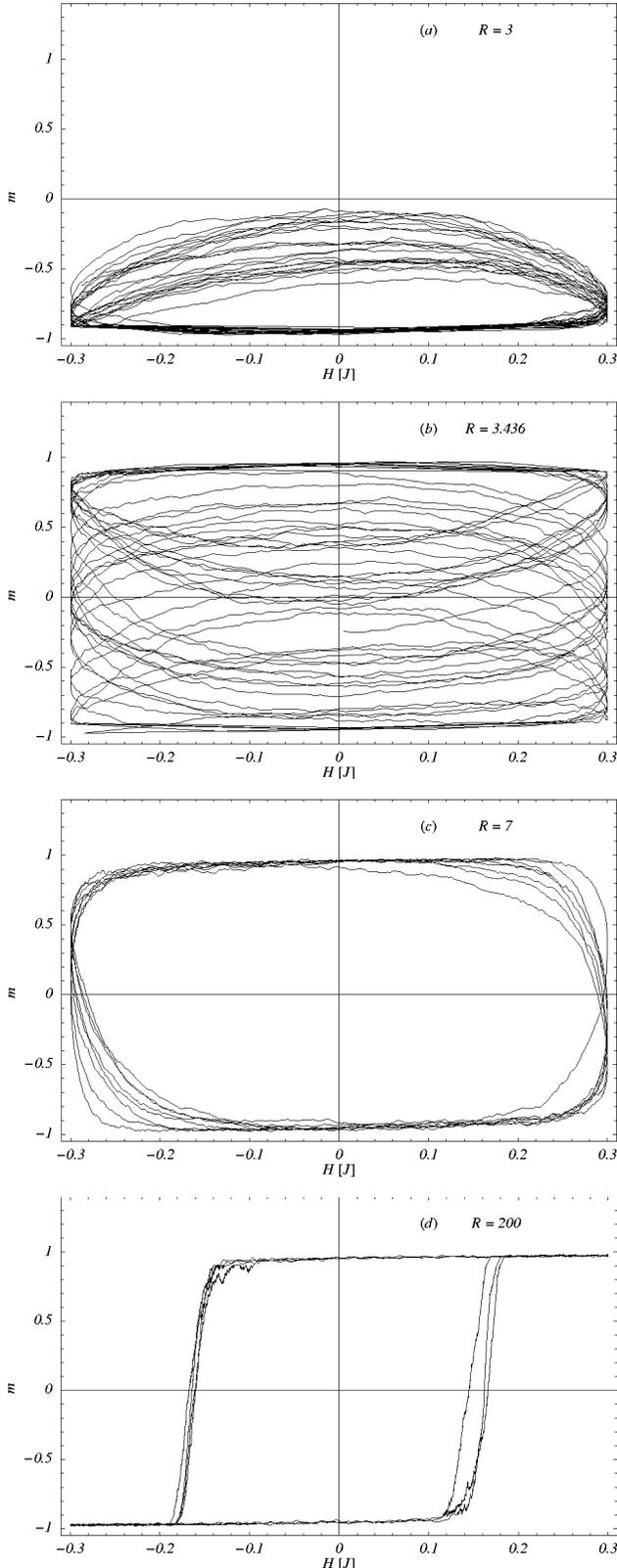

\centerline{\psfig{figure=mhMD_A_fig6a.epsi,width=3.25in,angle=270}}
\vspace{0.1in}
\centerline{\psfig{figure=mhMD_B_fig6b.epsi,width=3.25in,angle=270}}
\vspace{0.1in}
\centerline{\psfig{figure=mhMD_C_fig6c.epsi,width=3.25in,angle=270}}
\vspace{0.1in}
\centerline{\psfig{figure=mhMD_D_fig6d.epsi,width=3.25in,angle=270}}
\caption[Hysteresis loops in the MD regime]
{\label{fig_aLoops_md}
Representative hysteresis loops obtained from Monte Carlo simulation data
with system size $L \! = \! 64$ for
(a) $R \! = \! 3$,
(b) $R = 3.436 \approx R_{\rm cr}$,
(c) $R \! = \! 7$, and
(d) $R \! = \! 200$.
Each panel shows loops for the same
time intervals shown in the corresponding
time-series data in Fig.~\protect\ref{fig_mt_md}.
}
\end{figure}

The hysteresis loops depend on frequency and have qualitatively different
shapes above, near, and below the dynamic phase transition.
Figure \ref{fig_aLoops_md} shows plots of the magnetization vs 
field in the MD regime for four field frequencies.
The loops shown here are for the same frequencies as the time series shown
in Fig.~\ref{fig_mt_md}.
The loops shown in Fig.~\ref{fig_aLoops_md}(b) indicate the large fluctuations
in the average magnetization near the transition at
$R = 3.436 \approx R_{\rm cr}$.
As the frequency gets further away from the transition, the variation
in the loops becomes smaller.
The loops for a very low frequency, $R \! = \! 200$ 
in Fig.~\ref{fig_aLoops_md}(d),
show how the magnetization reverses sign early in each half-period of the
field, saturating close to the equilibrium magnetization at a field well
below $H_{0}$.
The low-frequency approximations for $A$
derived later in this section are applicable
to these squarish loops, in which the switching field is small
and the field variation is nearly linear.

\begin{figure}
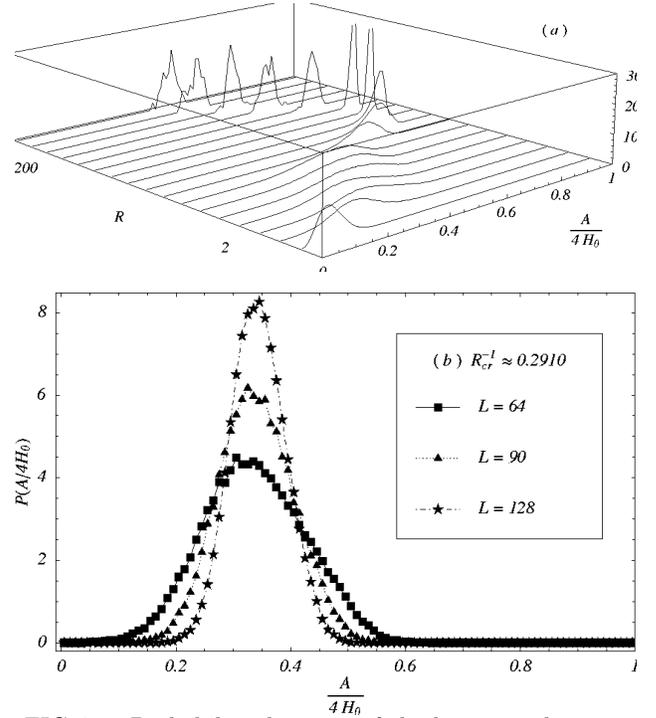

\centerline{\psfig{figure=a3dMD_L64_fig7a.epsi,width=3.25in,angle=270}}
\vspace{0.1in}
\centerline{\psfig{figure=aSame_r3.43_fig7b.epsi,width=3.25in,angle=270}}
\caption[Probability densities for the hysteresis-loop area (MD)]
{\label{fig_aDis_md}
Probability densities of the hysteresis-loop area,
$A = -\oint m(H) \ dH$.
The loop area has been normalized by the maximum possible loop
area, $4H_{0}$.
(a)
$L \! = \! 64$.
The values of the scaled period shown are
$R \! = \! 2$, $3$, $3.436 \approx R_{\rm cr}$,
$3.5$, $3.75$, $3.9$, $4$, $5$, $6$, $7$, $8.4$, $12$,
$25$, $50$, $80$, $140$, and $200$, so the $R$ axis is not linear. 
The two $R$ values explicitly marked indicate the directions
of increasing and decreasing frequency.
(b)
Distributions near the critical frequency shown for
$L \! = \! 64$, $90$, and 128.
The distributions are narrower away from $R_{\rm cr}$ but
look qualitatively similar.
}
\end{figure}

Probability densities for $A$
are shown in Fig.~\ref{fig_aDis_md}.
The loop area has been divided by the maximum possible loop area, $4H_{0}$.
Figure \ref{fig_aDis_md}(a) shows the probability densities
for $L \! = \! 64$ and several
values of $R$ between $R \! = \! 2$ and $R \! = \! 200$.
In contrast to the SD regime \cite{side98-SD},
the distributions are unimodal for {\it all} frequencies.
The roughness of the distributions at the lowest frequencies
is due to the lack of statistics because
the time series for these frequencies contain a smaller
total number of periods
than the data for the intermediate and high frequencies.
Figure \ref{fig_aDis_md}(b)
shows the probability densities for a frequency
near the transition for $L \! = \! 64$, $90$, and 128.
The position of the peak for each of these distributions appears to be
independent of system size.
However, the width of the distribution becomes smaller with increasing
system size.
This observation is consistent with the system size
dependence of the switching-field distribution
in field-reversal experiments \cite{rich95}.
For further
details on the $L$-dependence of the width near the transition,
see the end of Sec.~\ref{sec_q_md}.
The average loop area for a specific frequency is a quantity
often displayed
in experimental and numerical studies of hysteresis.
It can also be obtained in the present case, and we do so below.
The means of the distributions in Fig.~\ref{fig_aDis_md}(a) are shown
in Fig.~\ref{fig_aMean_md}, along with the average loop areas for
$L \! = \! 90$ and 128. 
The solid curve in Fig.~\ref{fig_aMean_md} is obtained by numerical
integration of Eq.~(\ref{eq_A_md}) using the values of $m(t)$
obtained in Sec.~\ref{sec_avrami}.

For any finite time series there is a sufficiently low frequency such that the
magnetization switches during every half-period of the field.
For very low frequencies the magnetization switches {\it before} the field
reaches its extreme value during every half-period.
In this frequency regime, $m(t)$ switches sign abruptly relative
to the length of a field period (see Fig.~\ref{fig_mt_md}(d)).
The switching time $t_{\rm s}$ may be found by solving 
$m(t_{\rm s}) \! = \! 0$.
Thus the equation for $t_{\rm s}$ becomes
 \begin{equation}
 \label{eq_ts1_md}
  \ln 2 = \Phi(t_{\rm s}) \ .
 \end{equation}
As Figs.~\ref{fig_mt_md}(d) and \ref{fig_aLoops_md}(d) show,
for sufficiently low frequencies
$m(t)$ switches sign early in the period, i.e. at a value $|H(t)| < H_{0}$.
This allows the approximation $H(t) \approx H_{0} \omega t$
to be used for the field, henceforth referred to as
``the linear approximation in the field.''
Using this approximation and the substitutions
$x = H_{0} \omega t$ and $x' = H_{0} \omega t_{\rm n}$
allows Eq.~(\ref{eq_Phi(t)_2}) to be rewritten as
 \begin{eqnarray}
  \Phi(t_{s})
   \nonumber
   & \approx &
    \frac{B(T) \Omega_{2} \nu^{2}}{4 H_{0}^{3} \omega^{3}}
    \int^{H_{\rm s}}_{0} x'^{3} (x^{2}-x'^{2})^{2}
    \exp \left [ -\frac{\Xi_{0}(T)}{x'} \right ] dx' \\
   \nonumber
   & = &
    \frac{B(T) \Omega_{2} \nu^{2}}{4 H_{0}^{3} \omega^{3}} \left \{
    x^{4} \int^{H_{\rm s}}_{0} x'^{3}
    \exp \left [ -\frac{\Xi_{0}(T)}{x'} \right ] dx' \right. \\
   \label{eq_phi_intermediate}
   & & -2 x^{2} \int^{H_{\rm s}}_{0} x'^{5}
    \exp \left [ -\frac{\Xi_{0}(T)}{x'} \right ] dx'  \\ \nonumber
   & & \left. + \int^{H_{\rm s}}_{0} x'^{7}
    \exp \left [ -\frac{\Xi_{0}(T)}{x'} \right ] dx'  \right \} \ .
 \end{eqnarray}
where $H_{\rm s} = H_{0} \omega t_{\rm s}$.
Together with Eq.~(\ref{eq_ts1_md}) this yields
 \begin{eqnarray}
  \ln 2
   \nonumber
   & = &
    \frac{B(T) \Omega_{2} \nu^{2}}{4 H_{0}^{3} \omega^{3}} \left \{
    H_{\rm s}^{4} \Xi_{0}^{4}(T)
    \Gamma\left[-4,-\frac{\Xi_{0}(T)}{H_{\rm s}} \right ] \right. \\
  \label{eq_linear1_md}
   & & 
   -2 H_{\rm s}^{2} \Xi_{0}^{6}(T)
   \Gamma\left[-6,-\frac{\Xi_{0}(T)}{H_{\rm s}} \right ] \\ \nonumber
  & & \left.
   +\Xi_{0}^{8}(T)
   \Gamma\left[-8,-\frac{\Xi_{0}(T)}{H_{\rm s}} \right ] \right \} \ ,
 \end{eqnarray}
where each of the integrals in Eq.~(\ref{eq_phi_intermediate})
has been rewritten as an incomplete gamma function
\cite{abra70} using the expression
 \begin{equation}
 \label{eq_gamma_md}
  \int^{x}_{0} u^{n} e^{-a/u} du
    = a^{n+1} \Gamma \left[ -(1+n), \frac{a}{x} \right ] \ .
 \end{equation}
For small $\omega$ the hysteresis loops are practically square, so the scaled
loop area in the low-frequency (LF) limit can be expressed as
 \begin{equation}
 \label{eq_loopLF_md}
 \nonumber
  \frac{\left < A \right >_{\rm LF}}{4H_{0}} 
  \approx m_{\rm eq}\frac{H_{\rm s}(\omega)}{H_{0}} \ .
 \end{equation}
The switching field $H_{\rm s}(\omega)$ 
is obtained by numerical solution of
Eq.~(\ref{eq_linear1_md}), and the result
for $\left < A \right >_{\rm LF}/4H_{0}$ is shown as the
dotted curve in Fig.~\ref{fig_aMeanLow_md}.
There is good agreement between the linear approximation calculation,
the numerical integration calculation (NI),
and the MC data for low frequencies.
The slight overestimate of the loop area by the linear approximation 
at low frequencies is
due to a systematic error in the way that the loop area is calculated from
the value of $H_{\rm s}(\omega)$.
The disagreement near the maximum in the loop area is
due to the breakdown of the linear approximation as the magnetization
begins to switch sign only at fields close to $H_{0}$.

The dash-dotted curve in Fig.~\ref{fig_aMeanLow_md} is the theoretical
low-frequency prediction for $L \! = \! 64$ in the SD regime.
This curve corresponds to the solid curve in
Fig.~11(b) of Ref.~\cite{side98-SD},
appropriately re-scaled so that it may be shown together with the MD results.
This re-scaling consists of using $H_{0} \! = \! 0.3J$ in the quantity
$A/4H_{0}$ and $\left < \tau(H_{0}) \right >$
for the MD regime in calculating the scaled frequency $1/R$.
The MC data point
at the very lowest frequency in Fig.~\ref{fig_aMeanLow_md}
($R \! = \! 5000$)
agrees with the SD, rather than the MD, theoretical
prediction, even though
the simulation was performed with $H_{0} \! = \! 0.3 J$.
This crossover from MD to SD hysteretic behavior is a completely
frequency-dependent effect.
As the frequency of the field decreases, the value of the switching
field, $H_{\rm s}(\omega)$ decreases as well.
For sufficiently low frequencies, $H_{\rm s}(\omega) < H_{\rm DSP}(L,T)$,
the system undergoes SD decay before the field becomes large
enough to produce MD decay.
In fact, the intersection of the SD and MD theoretical predictions occurs
at a value of $A$ which corresponds to a square loop with a
switching field of
$H_{\rm s}(\omega) \approx H_{\rm DSP}(L,T)$.
While $m(t)$ and $A$ do not depend on system size in the MD regime,
$A$ in the SD regime, and hence the location of the crossover, depends
on $L$ through the $L^{-d}$ dependence of the lifetime in the SD regime
(see Eqs.~(4.15) and (7.13) of Ref.~\cite{side98-SD}).

One can obtain an approximate analytic solution
for $\left < A \right >_{\rm LF}$ by taking the first three
terms in the asymptotic expansion \cite{abra70}
 \begin{equation}
 \label{eq_gammaExpansion_md}
  \Gamma(a,x) \sim x^{a-1} e^{-x}
    \Bigl [ 1 + \frac{a-1}{x} + \frac{(a-1)(a-2)}{x^2} + \ldots \Bigr ]
 \end{equation}
for large $x$.
Straightforward, but lengthy,
algebra gives the following asymptotic
expansion for Eq.~(\ref{eq_linear1_md})
 \begin{eqnarray}
 \label{eq_asymp1_md}
  \left(\frac{H_{\rm s}}{\Xi_{0}(T)} \right)^{11}
  \exp \left [ -\frac{H_{\rm s}}{\Xi_{0}(T)} \right ]
   & \approx &
   (D H_{0} \omega)^{3} \ ,
  \end{eqnarray}
where we define
 \begin{equation}
 \label{eq_D_def}
  D = \left(\frac{\ln 2}{2 B(T) \Omega_{2} \nu^{2} \Xi_{0}^{8}(T)}
       \right )^{\frac{1}{3}} \ .
 \end{equation}
With $B(T) \! = \! 0.02048 J^{-3} {\rm MCSS}^{-1}$ and the
values found in Table \ref{table_constants_md}
this gives $D \! = \! 17.905 J^{-1} {\rm MCSS}^{-1}$.
Ignoring the nonexponential prefactor in Eq.~(\ref{eq_asymp1_md}), solving
for the switching field $H_{\rm s}(\omega)$, and substituting it
into Eq.~(\ref{eq_loopLF_md}) results in the asymptotic, logarithmic
frequency dependence for the loop area,
 \begin{equation}
 \label{eq_asympFinal_md}
  \left <A \right >_{\rm LF} \approx
    \frac{4}{3} \Xi_{0}(T) \left [ -\ln (D H_{0} \omega) \right ]^{-1} \ .
 \end{equation}
As in the SD case \cite{side98-SD,side98-MMM}, from a log-log plot of the 
loop area versus frequency one can extract effective 
exponents from the data over nearly two decades in $1/R$. 
However, these effective exponents depend strongly on the
frequency range in which the fit is performed. 
Very small effective exponents 
that appear consistent with this picture were recently reported 
by Suen {\it et al.\/} for ultrathin Co films on Cu(001) \cite{SUEN99}. 

As for the SD regime, we stress that the asymptotic low-frequency behavior
described by Eq.~(\ref{eq_asympFinal_md})
would only be seen for {\it extremely} low frequencies, 
where the dotted and dashed curves in Fig.~\ref{fig_aMeanLow_md} 
come together. Therefore, due to the dependence of $H_{\rm DSP}$ on $L$
the crossover to SD droplet behavior will occur at frequencies
much larger than those for which this asymptotic expression is valid
even for large $L$. For a system size of $L \! = \! 10^9$ the ratio of
the full numerical solution to the asymptotic expression is
approximately two near the crossover frequency,
which for that system size would be near $1/R \approx 6 \times 10^{-9}$.

Relatively few studies have considered $B$, the
correlation of the magnetization with the external field.
Our theoretical derivations of $B$ are analogous to those for $A$.
The probability densities of $B$ in the MD regime 
are shown in Fig.~\ref{fig_bDis_md}.
Figure \ref{fig_bDis_md}(a) shows the probability densities for 
several values of $R$ between 2 and 200.
The source of the
roughness of the distributions at the lowest frequencies
is the same as in Fig.~\ref{fig_aDis_md}(a).
Figure \ref{fig_bDis_md}(b) shows the probability densities for a frequency
near the transition for $L \! = \! 64$, $90$, and 128.
The system size dependence
of the peak position and the width of the distributions is similar to that
for the loop-area distributions.
The means of the distributions in Fig.~\ref{fig_bDis_md}(a) are shown in
Fig.~\ref{fig_bMean_md}
along with the results for $L \! = \! 90$ and 128.
The solid curve is obtained by numerical integration of
Eq.~(\ref{eq_B_md}) similar to the procedure for $A$.
The agreement between the MC data and the theoretical prediction is
very good for low and high frequencies, but poor for intermediate frequencies,
where the MC data take on negative values.
This disagreement
is probably due to the procedure for calculating $m(t)$ during a half-period
in which shrinking droplets are present in the system 
(see Sec.~\ref{sec_avrami}).

As for the SD regime \cite{side98-SD},
the physical significance of the integrals $A$ and $B$
can be clarified by comparison with linear response theory.
One can easily find
that $A/(\pi H_0^2)$ and $2B/H_0^2$ correspond to the dissipative and 
reactive parts of the complex linear response function, respectively.
It is therefore natural to combine $A$ and $B$ into an analogous 
{\em nonlinear\/} response function, 
 \begin{equation}
 \label{eq_X_md}
  \hat{X}(H_0,T,\omega) 
= \frac{1}{H_0^2} \left[ 2B + \frac{i}{\pi} A \right] \;.
 \end{equation}
The frequency dependence of the norm of this response function, $|\hat{X}|$,
is shown in Fig.~\ref{fig_normResp_md}.
The maximum in $A$ (Fig.~\ref{fig_aMean_md})
and the sign change in $B$ (Fig.~\ref{fig_bMean_md}),
which occur 
close together in frequency, are characteristic 
behaviors of dissipative and reactive parts of a response function
near resonance.
The behavior of $\left < A \right >$ and $\left < B \right >$ in this
frequency range is very similar to that observed
in the SD regime \cite{side98-SD}, where
we associated it with stochastic resonance \cite{gamm98_review}.
Whether or not this name is appropriate for the low-frequency
synchronization of $m(t)$ with $H(t)$ in the MD regime, is
probably a matter of taste.
Although the overall fluctuations in $m(t)$ are small, the switching is
entirely driven by random nucleation on a microscopic scale.

\section{Dynamic phase transition}
\label{sec_q_md}

Although nonequilibrium phase transitions have been studied 
for over two decades, the understanding of their 
universality and scaling properties remains much weaker than for equilibrium 
critical phenomena.
Finite-size scaling is routinely
used to determine the location and critical exponents
of equilibrium, thermodynamic phase transitions from numerical
results for finite systems \cite{bind88,priv90}.
A true phase transition occurs only in the thermodynamic limit, i.e.\
as the system size approaches infinity while the energy and particle
densities are kept constant.
For simulations on finite systems, quantities such
as the susceptibility and specific heat
only display a bounded peak, rather than a divergence, as a function of
temperature or field.
However, for simulations performed on larger and larger lattices,
thermodynamic quantities approach the infinite-size limit.
It is this approach to the thermodynamic limit that is utilized to obtain
estimates of the critical exponents from various system sizes.

Second-order phase transitions in equilibrium systems are characterized
by a set of critical exponents,
each of which describes the behavior of a different quantity
at the critical point.
Three of these quantities and their associated exponents are \cite{huan87}:
the order parameter ($M \sim |\bar{t}|^{\beta} \ {\rm for} \ \bar{t} < 0$),
the susceptibility ($\chi \sim |\bar{t}|^{-\gamma}$),
and
the correlation length ($\xi \sim |\bar{t}|^{-\nu}$)
where $\sim$ denotes ``the singular behavior of.''
The field conjugate to the order parameter is understood to be zero, and
$\bar{t} \! = \! (T-T_{c})/T_{c}$ is the reduced temperature.
Finite-size scaling theory allows one to estimate
the critical exponents by measuring
the system size dependence of various quantities.
Combining the expression for the correlation length exponent
with the finite-size scaling assumption, $\xi(T_{c}(L)) \sim L$,
gives
 \begin{eqnarray}
 \label{eq_fss1}
  |T_{c}(L) - T_{c}|   & \propto & L^{-1/\nu} \ ,
 \end{eqnarray}
where
$T_{c}(L)$ can be defined as the location of the peak in the
susceptibility for a given $L$ \cite{bind88,priv90}.
When combined with the definitions for the critical exponents,
Eq.~(\ref{eq_fss1}) gives
 \begin{equation}
 \label{eq_fss2}
  \chi_{L}^{\rm peak} \propto L^{\gamma/\nu}
 \end{equation}
and
 \begin{equation}
 \label{eq_fss3}
  \left < |M|^{n} \right >_{L}
    \propto L^{-n(\beta/\nu)} \,
 \end{equation}
where
$\chi_{L}^{\rm peak}$ is the maximum value of the susceptibility
for a given $L$, and $\left < |M|^{n} \right >_{L}$ is the $n$th moment
of the norm of the order-parameter at $T_{c}$.

The period-averaged magnetization $Q$ has been proposed as a
``dynamic order parameter'' for systems exhibiting hysteresis
\cite{tome90,acha94_review,acha92-1,acha92-2,acha94,acha94-2,lo90,side98-PRL}.
Those studies of the Ising
model have suggested the existence of a dynamic phase
transition between
an ordered dynamic phase with
$\langle |Q| \rangle >0$ and
a disordered dynamic phase with
$\langle |Q| \rangle \approx 0$.
Figure \ref{fig_qDis_md} shows the probability densities of $Q$
in the MD regime for $L \! = \! 64$, $90$, and 128.
For each system size, as the frequency of $H(t)$ decreases, the
probability densities for $Q$ change from bimodal 
distributions with the two peaks each centered
around a nonzero value of $Q$, to unimodal distributions with
a peak around $Q \! = \! 0$.
Each of the $Q$ distributions shown in the three-dimensional
Fig.~\ref{fig_qDis_md}
is a histogram of $Q$ time-series values (see Fig.~\ref{fig_qt_md})
at a particular frequency.
These distributions suggest the presence of a second-order
dynamic phase transition.
In particular, the frequency dependence for these probability densities
is strikingly similar to the dependence on inverse
temperature for probability densities of
the equilibrium magnetization in the zero-field Ising model.
We therefore
identify the norm of the period-averaged magnetization, $|Q|$, as the
order parameter of the dynamic phase transition,
and we apply finite-size scaling theory in analogy
to the scaling theories used to quantify second-order phase transitions
in {\it equilibrium} systems.
Figure \ref{fig_qmean_md} shows the average norm of the
period-averaged magnetization, $\left < |Q| \right >$, for the
same system sizes as in Fig.~\ref{fig_qDis_md}.
This figure clearly suggests a DPT as the average
order parameter $\left < |Q| \right >$ changes from a
value near zero to a non-zero value.
Rather than a sudden change in the order parameter,
the transition region is ``smeared'' out due to finite-size effects.
The mean and standard deviation for $|Q|$ are system size dependent as well.
We quantify these finite-size effects below.

At a second-order phase transition there is a divergence in the susceptibility.
For equilibrium systems, the fluctuation-dissipation theorem relates the
susceptibility to fluctuations in the order parameter.
For the present system, it is not obvious what the field conjugate to $Q$
might be.
Therefore, we cannot measure the susceptibility directly.
However, we can calculate the variance in $|Q|$ as a function of
frequency and study its system size dependence.
We define $X$ as
 \begin{equation}
 \label{eq_suscep_md}
  X = L^{2} \ {\rm Var}(|Q|)
    = L^{2} \left [ \left < Q^{2} \right > 
      - \left < |Q| \right >^{2} \right ].
 \end{equation}
If the system were to obey a fluctuation-dissipation relation,
$X$ would be proportional to the susceptibility, and both
would scale with $L$ in the same manner.
Figure \ref{fig_suscep_md} shows $X$ vs $1/R$ for all three system sizes.
For all three values of $L$, $X$ displays a prominent peak
near the transition frequency, which increases
in height with increasing $L$,
while no finite-size effect is seen at lower and higher frequencies.
This finite-size effect in $X$ implies
the existence of a divergent length associated with the order-parameter
correlation function near the dynamic transition.
The observation that $P(|Q|)$ displays no peak near $|Q| \! = \! 0$ in the
ordered dynamic phase
is additional evidence for the second-order
(as opposed to first-order) nature of this transition \cite{eich96}.

The cumulant intersection method \cite{bind88,priv90} is useful for
determining the location of a second-order transition when the critical
exponents are not known.
In order to estimate the location of the transition
we define the ``dynamic'' fourth-order cumulant ratio
 \begin{equation}
 \label{eq_U_md}
  U_{L} = 1 - \frac{ \left < |Q|^{4} \right >_{L}}
                   {3 \left < |Q|^{2} \right >_{L}^{2}} \ ,
 \end{equation}
where $\left < |Q|^{n} \right > = \int_{0}^{\infty} |Q|^{n} P(|Q|) d|Q|$.
Figure \ref{fig_cumulant_md} shows $U_{L}$ vs $1/R$ for the same system
sizes shown in Fig.~\ref{fig_suscep_md}.
Above the transition frequency, in the 
$\left < |Q| \right > > 0$ ordered dynamic phase,
$U_{L}$ approaches $2/3$,
corresponding to two narrow peaks centered at $\pm \left < |Q| \right >$.
Below the transition frequency, in the
$\left < |Q| \right > \approx 0$ disordered dynamic phase,
$U_{L}$ approaches $0$,
corresponding to a Gaussian centered at zero.
At the transition, the cumulant
should have a non-trivial fixed value, $U^{*}$.
Therefore, the location of the cumulant intersection
gives an estimate of
the transition frequency without foreknowledge of
the critical exponents.
Due to the large spacing of our data and possible
correction-to-scaling
effects, we cannot identify a unique intersection point.
We estimate the location of the intersection by the
crossing for the two largest system sizes near
$R_{\rm cr}^{-1} \approx 0.2910$ ($R_{\rm cr} \approx 3.436$)
with $U_{L} = U^{*} \approx 0.61$.
This is close to an extremely precise transfer
matrix calculation of $U^{*} \! = \! 0.6106901(5)$ \cite{blot93}
as well as MC estimates \cite{mon97} for the two-dimensional Ising model. 
However, as recently pointed out by Luijten {\em et al.\/} \cite{LUIJ98}, 
the value of the cumulant intersection should not be taken too seriously 
unless a sufficient range of system sizes is available. 

With our estimate for the transition frequency, we can now
approximate the critical
exponents $\beta$, $\gamma$, and $\nu$ characterizing
the transition by using
Eqs.~(\ref{eq_fss1})-(\ref{eq_fss3}), replacing
$T_{c}$ by $R_{c}$,
$\chi^{\rm peak}_{L}$ by $X^{\rm peak}_{L}$, and
$\left<|M|^{n}\right>_{L}$ by $\left<|Q|^{n}\right>_{L}$.
To extract exponent estimates using these relations, we
use a method sometimes referred to as
``phenomenological renormalization'' \cite{priv90} of the MC data.
This method consists of estimating an exponent by using two system sizes,
$bL$ and $L$.
The following example is a derivation of an exponent estimate for 
$\beta/\nu$.
 \begin{eqnarray}
  \nonumber
  \frac{\left < |Q|^{n} \right >_{bL}}{\left < |Q|^{n} \right >_{L}}
  & \propto &
  \frac{(bL)^{-n(\beta/\nu)}}{L^{-n(\beta/\nu)}} = b^{-n(\beta/\nu)} \ ,
 \end{eqnarray}
which yields
 \begin{eqnarray}
  \label{eq_lnb_derivation1}
  -\left.\ln \left[ \frac{\left < |Q|^{n} \right >_{bL}}
                   {\left < |Q|^{n} \right >_{L}} \right ] \right / \ln b
  & = & n \left (\frac{\beta}{\nu} \right ) + O\left (1/\ln b \right)   \ .
 \end{eqnarray}
Similar relations can be found for the other exponent ratios,
 \begin{eqnarray}
  \label{eq_lnb_derivation2}
  \left.\ln \left[ \frac{X^{\rm peak}_{bL}}
                   {X^{\rm peak}_{L}} \right ] \right / \ln b
  & = & \frac{\gamma}{\nu} + O\left (1/\ln b \right) \\
  \label{eq_lnb_derivation3}
  \left.\ln \left[ \frac{|R_{c}(bL)-R_{c}|}
                   {|R_{c}(L)-R_{c}|} \right ] \right / \ln b
  & = &  \frac{1}{\nu} + O\left (1/\ln b \right)  \ .
 \end{eqnarray}
In the large-system limit
these exponent estimates will be linear when plotted vs $(\ln b)^{-1}$.
Then, one can extrapolate to the infinite
size limit by performing a linear fit of the data to find the intercept
at $(\ln b)^{-1} \! = \! 0$.
Simulations with larger system sizes would be computationally
prohibitive, and smaller system sizes would no longer be in the
MD regime. With data for only three system sizes, the exponent
estimates obtained using the two largest system sizes are 
easily shown to be identical to those obtained using the extrapolation
procedure above. We calculate two sets of estimates for $\beta/\nu$,
one using the scaling relation for the second moments
of the order-parameter
distribution ($n \! = \! 2$) and the other using ($n \! = \! 4$),
obtaining $(\beta / \nu)_{n=2} \approx 0.111$
and $(\beta / \nu)_{n=4} \approx 0.113$.
Our estimates for the other exponents are
$\gamma / \nu \approx 1.84$ and $\nu \approx 1.1$.
Also, we obtained an independent estimate for the exponent $\nu$ by
measuring the finite-size effects in the location of the high-frequency
zero-crossing in $\left < B \right >$.
The estimate obtained is $\nu \approx 1.09$, in good agreement with that 
obtained from the location of $X^{\rm peak}$.
Our results are close to the two-dimensional
Ising values for the analogous exponent ratios
($\beta/\nu \! = \! 1/8 \! = \! 0.125$, 
$\gamma/\nu \! = \! 7/4 \! = \! 1.75$, $\nu \! = \! 1$).
Given the accuracy of our data however,
our exponent estimates are also not inconsistent with the universality class of
two-dimensional, random percolation
($\beta/\nu \! = \! 5/48  \! \approx  \! 0.104$, 
$\gamma/\nu \! = \! 43/24  \! \approx  \! 1.79$,
$\nu \! = \! 4/3  \! \approx  \! 1.33$).
Combining the exponent estimates we find
 \begin{equation}
 \label{eq_hyperscaling}
  2(\beta/\nu)+(\gamma/\nu) \approx 2.06 \approx d \ .
 \end{equation}
This relation between the measured exponent ratios indicates the 
consistency of our scaling procedures and thus strengthens our belief that 
the dynamic transition is a genuine, continuous phase transition. 
If the divergent length is indeed the correlation length that describes 
the order-parameter correlation function, then Eq.~(\ref{eq_hyperscaling}) 
is a hyperscaling relation. The DPT critical point then should be a 
nontrivial fixed point in the renormalization group sense. 
Based on the evidence presented in the following paragraph, we believe 
this is the case. 

One should also consider the possibility that hyperscaling is 
violated and the DPT represents a mean-field critical point. The divergent 
length would then be the ``thermal length'' \cite{bind88,LUIJ98}, 
whose divergence is governed by the exponent $(2 \beta + \gamma) / d$. 
This would then be the exponent we have called ``$\nu$,'' and 
Eq.~(\ref{eq_hyperscaling}) would hold exactly as a tautology. 
However, our estimates for $\beta$ and $\gamma$ are far from those of a 
mean-field $\phi^4$ model ($\beta \! = \! 1/2$ and $\gamma \! = \! 1$). 
Likewise, our estimated cumulant crossing, 
$U^\ast \! \approx \! 0.61$, is far from the expected mean-field value, 
$U^\ast = 1 - \Gamma^4(1/4)/(24 \pi^2) \approx 0.27$ \cite{LUIJ98,BREZ85}. 
Furthermore, if the phase transition were to be mean-field in this 
two-dimensional system, it should have to be induced by some effective 
long-range interaction, which then should have the same effect in one 
dimension. However, exploratory MC simulations indicate 
that the one-dimensional Ising model in an oscillating field does not 
have a dynamic ordered phase \cite{KORN98}. 
The evidence summarized in this paragraph makes it extremely unlikely that 
hyperscaling is violated by the DPT. 

The consistency of the estimates of $\nu$ from the positions
of $X^{\rm peak}$ and the high-frequency zero-crossing of
$\left < B \right >$
indicates that this zero occurs at the DPT.
The two zeros of $\left < B \right >$ are clearly separated in frequency, 
and our finite-size scaling results indicate that they remain so as 
$L \rightarrow \infty$. 
The low-frequency zero is associated with the maximum in  
$\left < A \right >$.
These observations enable us to answer the question recently raised
by Acharyya \cite{acha98} of whether the DPT corresponds to the maximum in
$\left < A \right >$.
It does not.

To further illustrate the nature of the dynamic phase transition,
the finite-size effects in the distributions for the norm of the
order-parameter, $|Q|$, are shown in Fig.~\ref{fig_symmQdis_md}.
These probability densities for $|Q|$ are the symmetrized versions
of selected distributions shown in Fig.~\ref{fig_qDis_md}.
Distributions in the ordered
dynamic phase region, i.e.\ above the transition frequency,
should move toward a constant, nonzero value
of $|Q|$ and become narrower with increasing $L$.
This is seen in Fig.~\ref{fig_symmQdis_md}(a).
For this frequency, the distribution of $Q$
for $L \! = \! 90$ is highly asymmetric about zero
and for $L \! = \! 128$ the distribution is unimodal.
This gradual loss of symmetry with increasing $L$ is
due to the finite length of the simulation time series,
but it does not adversely affect our ability to analyze $P(|Q|)$.
The distributions in Fig.~\ref{fig_symmQdis_md}(b) are
in the disordered dynamic phase
region, i.e.\ at a frequency slightly
below the transition frequency.
Due to finite-size effects however,
the distributions for $L \! = \! 64$ and $90$ appear to be centered
about nonzero values of $|Q|$.
The distributions in Fig.~\ref{fig_symmQdis_md}(c) are
near the transition and should scale with system size $L$.
We assume that the mean of the order parameter scales
with $L$ and define the scaling variable 
$\tilde{Q} = L^{\beta / \nu}|Q|$.
Hence, the scaled probability density for $|Q|$ is given by
 \begin{equation}
 \label{eq_scaling_dist}
  \tilde{P}_{L}(\tilde{Q}) = L^{-\beta / \nu} P(|Q|) \ ,
 \end{equation}
where the prefactor $L^{-\beta / \nu}$ ensures conservation of probability. 
Figure \ref{fig_scaledQdis_md} shows this
scaled probability density.
The peak positions scale fairly well, the peak heights less so.
This could be due to the following reasons.
The frequency might be sufficiently far
from the transition that single-parameter scaling is not adequate,
and there might be corrections to the finite-size scaling
that are large for these relatively small system sizes.
Also, the lack of scaling for the peak heights could
be due to the asymmetry in $P(Q)$ near the transition.

The results in this section clearly show that the
statistical properties of the order parameter $Q$
exhibit finite-size scaling, and that
scaling techniques developed for estimating the
critical exponents for second-order phase transitions
in equilibrium systems apparently can be
successfully applied to estimate the exponents associated with
the dynamic phase transition.
While these scaling relations are concerned with $|Q|$,
it is worth mentioning that one may also measure the fluctuations in the
other two quantities measured, $A$ and $B$.
Figure \ref{fig_suscepAB_md} shows the fluctuations for $A$ and $B$,
defined in analogy to the order-parameter fluctuation $X$.
The fluctuations in $A$ seem to show slight finite-size effects,
as the peak positions appear to be approaching $R_{\rm cr}^{-1}$
with increasing $L$.
One might speculate that this could indicate that $A$ is coupled to
energy fluctuations which are logarithmically divergent
as they are for the
two-dimensional Ising model in equilibrium.
The fluctuations in $B$ show no significant
finite-size effects.

\section{Discussion}
\label{sec_md_conclusion}

The mechanism by which a metastable phase decays depends sensitively on
the system size, the temperature, and the strength of the applied field.
For large systems and
moderately strong fields, the decay proceeds through
the nucleation and growth of {\it many} droplets of overturned spins
in different parts of the system.
This regime has been termed the multi-droplet (MD) regime.
In this regime the magnetization response in static field is
described by the KJMA approximation (Avrami's law),
which assumes the presence
of many noninteracting, overlapping droplets.
Theoretical predictions
by a generalization of the KJMA approximation, in
which a time-dependent nucleation
rate and droplet interface velocity are used, 
agree well with simulations
for quantities like the average
hysteresis-loop area and correlation, 
especially at low driving frequencies. 
The time dependence is included in the theory
by replacing the
constant field $H$ by $H(t) \! = \! -H_{0} \sin(\omega t)$.
This central idea provides the analytic framework for our theoretical
descriptions of quantities measured from the MC simulations
in the MD regime.
The theoretical calculations and the MC data agree very well, especially
considering that only one adjustable parameter is needed,
which was measured from a particular hysteresis simulation ($R \! = \! 200$).
All of the other constants used are either known from droplet theory or
were measured for MC simulations of
field reversal in kinetic Ising models.
To the best of our knowledge,
the work reported here is the first which explicitly
considers hysteresis for the Ising model in the MD regime.

We compute the power spectral densities from the simulated time series
and qualitatively explain various features of the spectra in the full
frequency range from the lowest observable frequencies to the rapid
fluctuations due to thermal noise.
For low field frequencies, the system is in the disordered dynamic phase,
and the time series contain no large fluctuations.
Consequently, the PSDs are flat at frequencies below the fundamental peak
at the frequency of the field. The significant power density in
the low-frequency portion of the PSDs corresponds to
the long-time behavior in the filtered time series for $Q$.
For high field frequencies, the system is in the ordered dynamic phase,
and the time series display long-time behavior as the system
switches between thermodynamic phases.
This corresponds to large power density
in the low-frequency portion of the PSDs.
Near the dynamic phase
transition the PSDs exhibit similar behavior in the low-frequency
part of the spectrum.
Due to insufficiently long time series we are
unable to resolve any quantitative difference between these PSDs and those
in the ordered dynamic phase region.

\begin{figure}
\centerline{\psfig{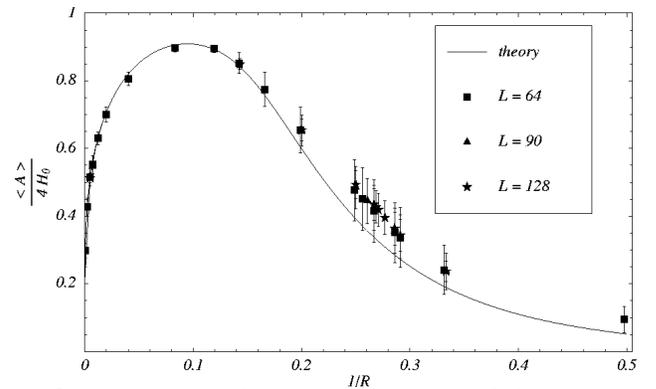}}
\caption[Mean loop area vs frequency]
{\label{fig_aMean_md}
Mean and standard deviation of the loop-area distributions vs the
scaled frequency $1/R$.
The data points are the means of the distributions shown in
Fig.~\protect\ref{fig_aDis_md} for $L \! = \! 64$ along with
the means of the corresponding
distributions for $L \! = \! 90$ and 128.
The vertical bars are {\it not} error bars, but give the standard deviations
of those distributions.
The solid curve comes from numerical integration of
Eq.~(\protect\ref{eq_A_md}) using the values of
$m(t)$ from Sec.~\protect\ref{sec_avrami}.
The single free parameter, $B(T)$,
is adjusted so the theoretical prediction agrees with
the data point at $1/R \! = \! 0.005$ as described in the text.
}
\end{figure}

\begin{figure}
\centerline{\psfig{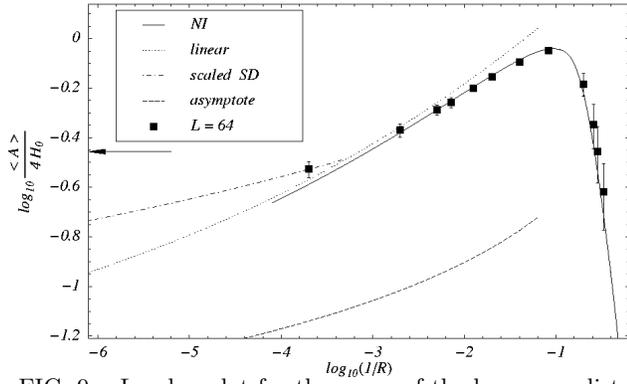}}
\caption[Log-log plot of mean loop area vs frequency]
{\label{fig_aMeanLow_md}
Log-log plot for the mean of the loop-area distributions vs the
scaled frequency $1/R$.
The vertical bars are {\it not} error bars, but give the standard deviations
of those distributions.
The data points are the same as those used for $L \! = \! 64$ in
Fig.~\protect\ref{fig_aMean_md} for the lowest frequencies.
The solid curve (NI)
is the same as the solid
curve in Fig.~\protect\ref{fig_aMean_md}.
(Due to numerical difficulties, this calculation was not extended to 
lower frequencies than those shown.)
The dotted curve results from numerical solution of
the linear approximation, Eq.~(\ref{eq_linear1_md}).
The dash-dotted curve is the prediction for the loop area
in the SD regime for $L \! = \! 64$,
scaled so that it may be plotted along with the MD results. 
The dashed curve is the asymptotic, logarithmic frequency
dependence for the MD loop area, Eq.~(\protect\ref{eq_asympFinal_md}).
This asymptotic result approaches the full solution only for frequencies
that are lower than the crossover to the SD solution, even for very
large $L$.
The arrow indicates the area of a hysteresis
loop with $H_{s} \approx H_{\rm DSP}(L \! = \! 64)$.
}
\end{figure}

\begin{figure}
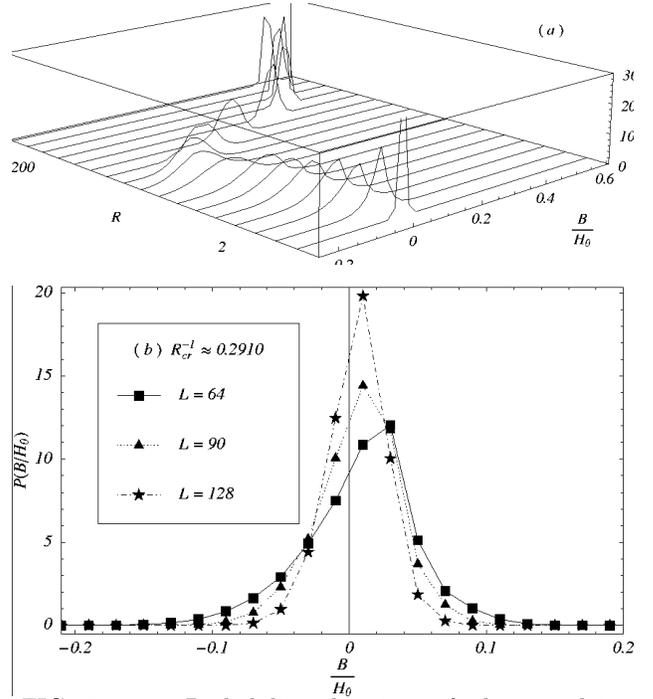

\centerline{\psfig{figure=b3dMD_L64_fig10a.epsi,width=3.25in,angle=270}}
\vspace{0.1in}
\centerline{\psfig{figure=bSame_r3.43_fig10b.epsi,width=3.25in,angle=270}}
\caption[Probability densities for the correlation (MD)]
{\label{fig_bDis_md}
Probability densities of the correlation,
$B = (\omega/2\pi)\oint m(t) \ H(t) \ dt$.
(a)
$L \! = \! 64$.
The values of $R$ shown are the same as in
Fig.~\protect\ref{fig_aDis_md}.
(b)
Distributions near the critical frequency shown for
$L \! = \! 64$, $90$, and $128$.
The lines are guides to the eye. 
The distributions look qualitatively similar, even away from $R_{\rm cr}$.
}
\end{figure}

\begin{figure}
\centerline{\psfig{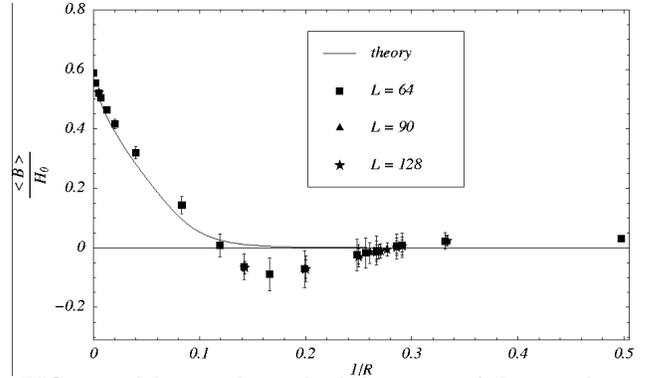}}
\caption[Mean correlation vs frequency (MD)]
{\label{fig_bMean_md}
Mean and standard deviation of the correlation distributions vs the
scaled frequency $1/R$.
The data points are the means of the distributions shown in
Fig.~\protect\ref{fig_bDis_md}(a)
for $L \! = \! 64$,
along with the corresponding results for $L \! = \! 90$ and $128$.
The vertical bars are {\it not} error bars, but give the standard deviations
of the distributions.
The solid curve comes from a theoretical calculation analogous
to that in Fig.~\protect\ref{fig_aMean_md}.
}
\end{figure}

\begin{figure}
\centerline{\psfig{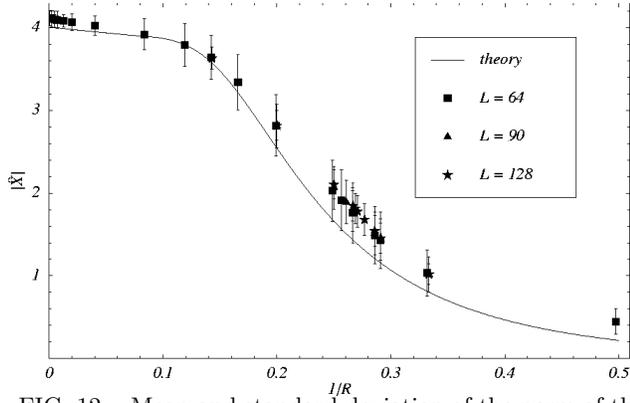}}
\caption{
\label{fig_normResp_md}
Mean and standard deviation of the
norm of the response function, $|\hat{X}|$, vs the scaled frequency $1/R$.
The data points are obtained using the loop-area and correlation data
in Fig.~\protect\ref{fig_aMean_md} and Fig.~\protect\ref{fig_bMean_md}.
The solid curve is obtained using the theoretical values for
$\left < A \right >$ and $\left < B \right >$ in those figures.
}
\end{figure}

\begin{figure}
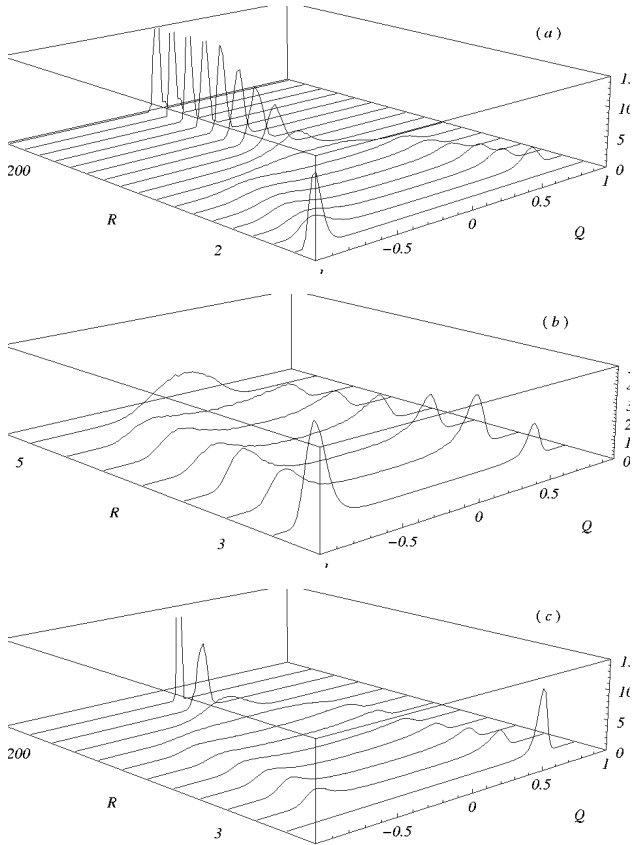

\centerline{\psfig{figure=q3dMD_L64_fig13a.epsi,width=3.25in,angle=270}}
\vspace{0.1in}
\centerline{\psfig{figure=q3dMD_L90_fig13b.epsi,width=3.25in,angle=270}}
\vspace{0.1in}
\centerline{\psfig{figure=q3dMD_L128_fig13c.epsi,width=3.25in,angle=270}}
\caption[Probability densities for the period-averaged magnetization (MD)]
{\label{fig_qDis_md}
Probability densities of the period-averaged magnetization,
$Q = (\omega/2\pi) \oint m(t) \ dt$.
(a)
$L \! = \! 64$.
The values of $R$ shown are the same as in
Fig.~\protect\ref{fig_aDis_md}.
(b)
$L \! = \! 90$.
The values of the scaled period shown are
$R \! = \! 3$, $3.436 \approx R_{\rm cr}$,
$3.5$, $3.75$, $3.835$, $4$, and $5$.
(c)
$L \! = \! 128$.
The values of the scaled period shown are
$R \! = \! 3$, $3.436 \approx R_{\rm cr}$, $3.5$, $3.612$, $3.693$, $3.721$,
$3.75$, $4$, $5$, $7$, and $200$.
The unimodal distributions for the smallest $R$ value in (a) and (c)
are due to the finite simulation time.
}
\end{figure}

\begin{figure}
\centerline{\psfig{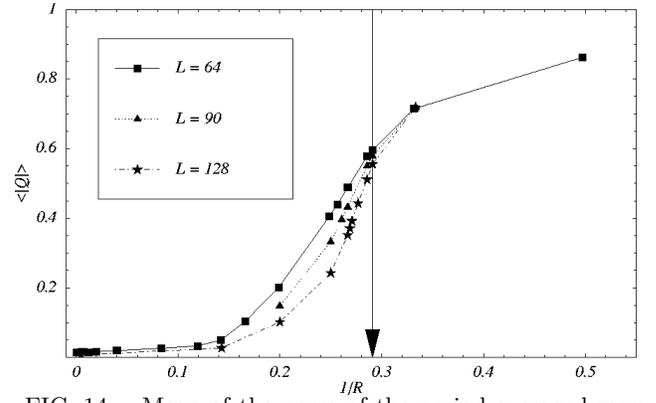}}
\caption[Mean of $|Q|$]
{\label{fig_qmean_md}
Mean of the norm of the period-averaged magnetization
vs the scaled frequency $1/R$.
The finite-size effects are clearly seen for
frequencies in the
neighborhood of the dynamic phase transition.
The arrow indicates the approximate value of the critical frequency
$1/R_{\rm cr}$.
Lines connecting the data points are guides to the eye. 
}
\end{figure}

\begin{figure}
\centerline{\psfig{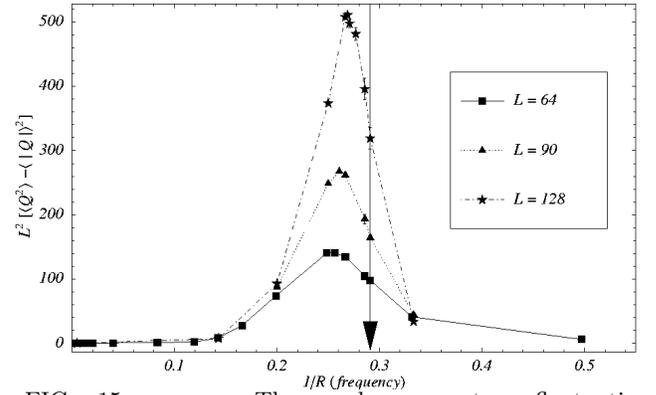}}
\caption[Variance in period-averaged magnetization]
{\label{fig_suscep_md}
The order-parameter fluctuation 
$X \! = \! L^{2} \ {\rm Var}(|Q|)$ vs the scaled frequency $1/R$.
The ``disordered dynamic phase,''
($\langle |Q| \rangle \! \approx \! 0$), lies on the
low-frequency side of the peaks.
The ``ordered dynamic phase,'' ($\langle |Q| \rangle > 0$),
lies on the high-frequency side.
The statistical error bars are estimated by partitioning
the data into ten blocks.
Error bars smaller than the symbol sizes are not shown.
The arrow indicates the approximate value of the critical frequency
$1/R_{\rm cr}$.
Lines connecting the data points are guides to the eye.
}
\end{figure}

\begin{figure}
\centerline{\psfig{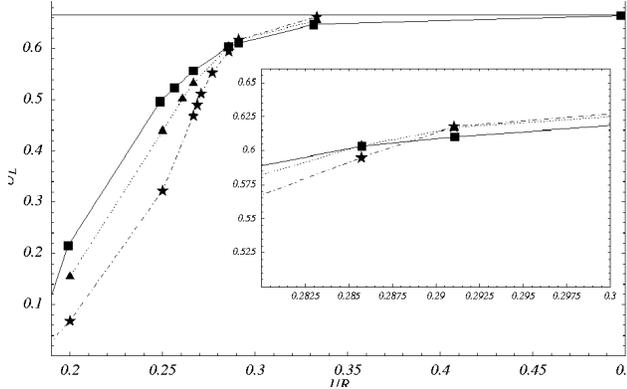}}
\caption[Fourth-order cumulant ratios]
{\label{fig_cumulant_md}
Fourth-order cumulant ratio $U_{L}$ vs scaled frequency $1/R$, 
for $L \! = \! 64$, 90, and 128.
We use the same symbols as in Fig.~\protect\ref{fig_suscep_md}.
The horizontal line marks $U_{L} \! = \! 2/3$.
Lines connecting the data points are guides to the eye.
Inset: area close to the cumulant crossing
at $1/R_{\rm cr} \approx 0.2910$.
}
\end{figure}

\begin{figure}
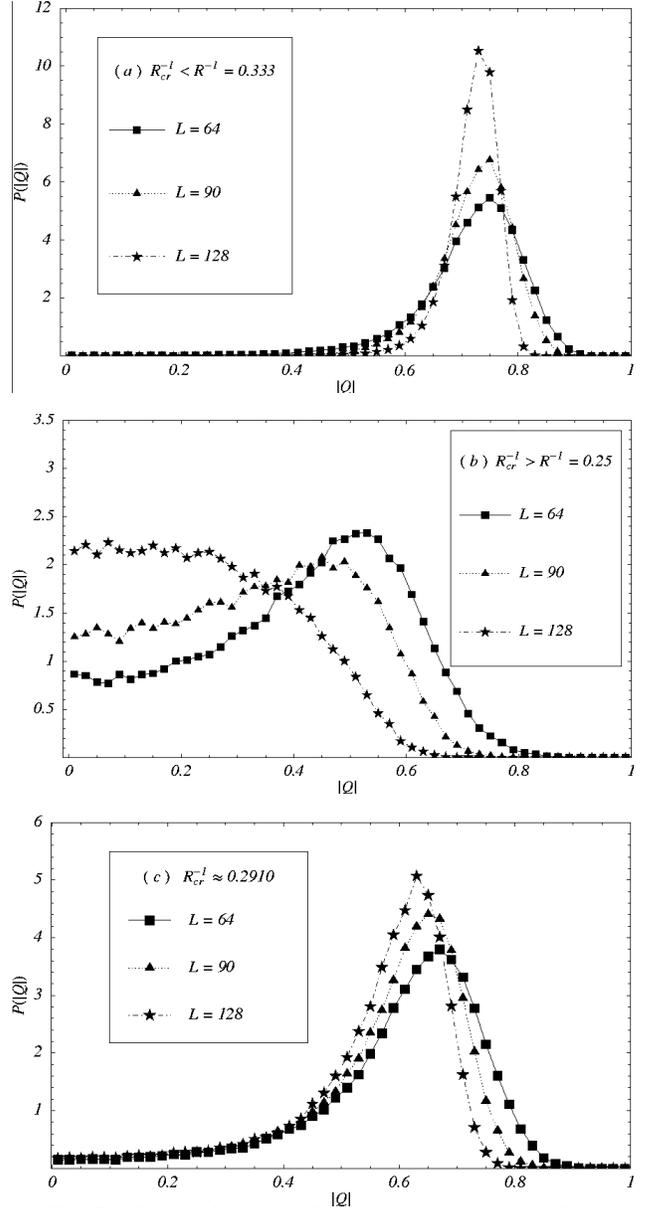

\centerline{\psfig{figure=symmDis_r3_fig17a.epsi,width=3.25in,angle=270}}
\vspace{0.1in}
\centerline{\psfig{figure=symmDis_r4_fig17b.epsi,width=3.25in,angle=270}}
\vspace{0.1in}
\centerline{\psfig{figure=symmDis_r3.43_fig17c.epsi,width=3.25in,angle=270}}
\caption[Symmetrized period-averaged magnetization distributions (MD)]
{\label{fig_symmQdis_md}
Probability distributions for the norm of the period-averaged magnetization
$|Q|$ for a frequency
(a) above the transition, $1/R \! = \! 0.333$,
(b) below the transition, $1/R \! = \! 0.25$,
and (c) near the transition $1/R \! = \! 0.2910$.
Lines connecting the data points are guides to the eye.
}
\end{figure}

\begin{figure}
\centerline{\psfig{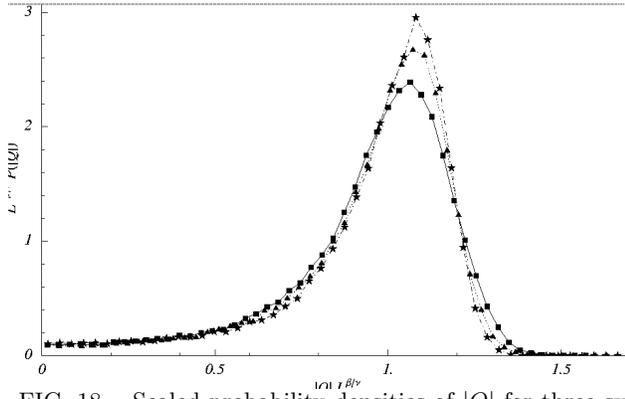}}
\caption[Period-averaged magnetization distribution at the transition]
{\label{fig_scaledQdis_md}
Scaled probability densities of $|Q|$ for three system sizes.
The same symbols are used as in Fig.~\protect\ref{fig_suscep_md}.
The scaling function is
$L^{-\beta / \nu} P(|Q|)$ vs $L^{\beta / \nu} |Q|$, and
the value of the scaling exponent used is
$(\beta/\nu)_{n=2} \approx 0.11$.
The scaled
frequency of the field is $1/R = 0.2910 \approx 1/R_{\rm cr}$.
Lines connecting the data points are guides to the eye.
}
\end{figure}

\begin{figure}
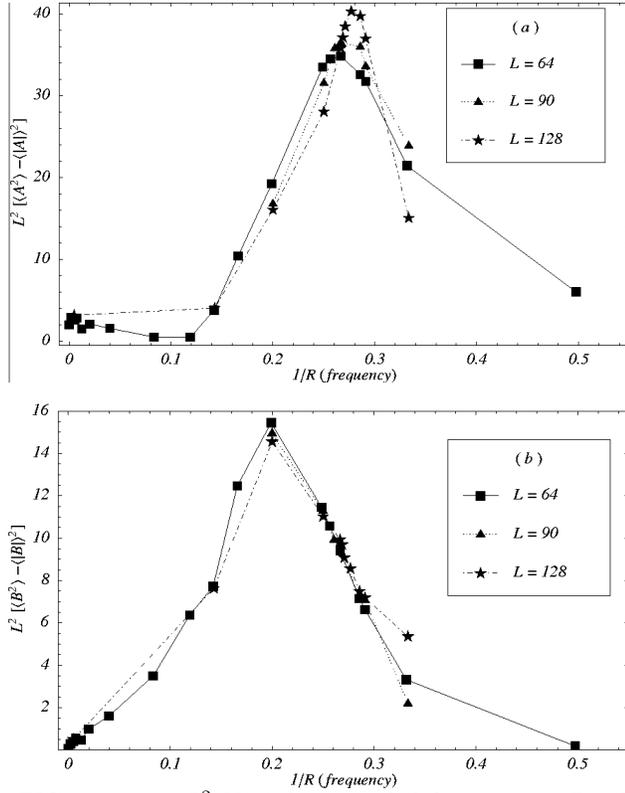

\centerline{\psfig{figure=susAplt_fig19a.epsi,width=3.25in,angle=270}}
\vspace{0.1in}
\centerline{\psfig{figure=susBplt_fig19b.epsi,width=3.25in,angle=270}}
\caption[Variance in loop area and correlation]
{\label{fig_suscepAB_md}
(a) $L^{2}$ Var$(|A|)$ vs scaled frequency $1/R$.
(b) $L^{2}$ Var$(|B|)$ vs scaled frequency $1/R$.
In both parts, the lines connecting the data points are guides to the eye.
}
\end{figure}

We also calculate the hysteresis-loop area $A$ and the correlation $B$
for a wide range of frequencies.
Because of its role as a measure of the energy dissipation in the system,
$A$ is a quantity of particular experimental significance.
For all frequencies, the loop-area and correlation distributions
are unimodal due to the almost
deterministic magnetization response in the MD regime.
Our theoretical predictions for the frequency dependence of
$\left < A \right >$ and $\left < B \right >$ use
the time-dependent extension
of Avrami's law to calculate $m(t)$, from
which the loop area and correlation are calculated explicitly.
The assumption is that the $m(t)$ values calculated for a {\it single}
period accurately describe the average values of $A$ and $B$ over a long
simulation time series.
This assumption should be expected to break down most significantly for
frequencies near the
dynamic phase transition, where the fluctuations in the magnetization
response are largest.
This is clearly seen in Figs.~\ref{fig_aMean_md} and ~\ref{fig_bMean_md},
where the least satisfactory
agreement between the theory and the MC data occurs for frequencies
near the dynamic phase transition.
For the low-frequency regime we obtain an analytic expression for
$\left < A \right >$.
Our theoretical calculation agrees well with our MC results and predicts
an {\it extremely} slow crossover to a logarithmic dependence of the
loop area on $H_{0}\omega$.
The switching dynamics is dominated by nucleation and indicates no
overall power-law dependence for the loop area on field amplitude
and/or frequency, in contrast to what has been claimed in other
simulational and experimental studies.
We emphasize that numerical analysis of data generated by
our analytic solution, even over two or three frequency decades,
could easily lead to the conclusion that the data were taken from a power law.
Our simulations reveal that for frequencies {\it far} higher than those
at which the asymptotic logarithmic dependence would be observable,
a system-size dependent crossover from MD to SD behavior occurs.
This novel frequency dependence for $A$ is a consequence of the field
dependence of the SD and MD decay mechanisms. 
As the frequency of the field becomes sufficiently small, the system is subject
to fields smaller than $H_{\rm DSP}(T,L)$
for a sufficiently long time so that SD decay usually occurs
before the field becomes large enough for MD decay to happen.

The period-averaged magnetization $Q$ has been proposed as an
order parameter
associated with
the dynamic phase transition (DPT) in kinetic Ising models.
The DPT is a {\it nonequilibrium} phase transition which occurs due to an
explicit time-dependence in the Hamiltonian, rather than the dynamical rules
governing the system.
The probability densities that we obtain
for $Q$ clearly show that the system changes from an
ordered dynamic phase with nonzero $\left < |Q| \right >$ to a
disordered dynamic phase
with $\left < |Q| \right > \approx 0$ as the field
moves from high to low frequencies.
To distinguish this frequency-dependent change in
$\left < |Q| \right >$ as a true second-order
phase transition rather than merely
a simple bifurcation, we measure the finite-size effects at the DPT
and apply finite-size scaling (FSS)
techniques analogous to those used to measure the critical exponents
which characterize {\it equilibrium} second-order transitions.
The measured exponents
($\beta / \nu \approx 0.11$,
$\gamma / \nu \approx 1.84$, and $\nu \approx 1.1$)
are close to both the two-dimensional Ising and random percolation
values, and they represent strong evidence that hyperscaling is obeyed.
Our success in applying FSS techniques borrowed from
the theory of equilibrium
second-order phase transitions to this
nonstationary nonequilibrium problem
suggests the possibility of
mapping other suitably defined quantities for this system
to thermodynamic entities, such as the field conjugate to the order parameter,
the specific heat, and the correlation length.
Such a nonequilibrium thermodynamic theory for steady states as been attempted
by Paniconi and Oono \cite{oono97}.

In future work we plan to
analyze longer simulations on larger system sizes
to more accurately determine the values of the exponents and the location
of the DPT. 
This includes measuring the possible finite-size effects in the
probability distributions of the energy \cite{acha97-2}, which might be 
related to the finite-size effects seen in the fluctuations in $A$.
If a fluctuation-dissipation theorem could be shown for this system,
the order-parameter and energy fluctuations could
be related to a nonequilibrium susceptibility and specific heat
respectively.
Another important question left to future study is
if and to what extent
the exponents depend on the temperature and field amplitude.
While the critical frequency will almost certainly depend on amplitude
and temperature, the critical exponents would most likely
not if the DPT indeed represents
a new ``dynamic universality class.''

Finally, we note that the quantities that we have analyzed numerically could 
all be measured in experiments on hysteresis in a variety of
systems and analyzed by methods essentially identical to our analysis of 
the MC data.

\acknowledgments{

Thanks to
M.\ Acharyya,
P.~D.\ Beale, 
G.\ Brown,
W.\ Janke,
W.\ Klein,
M.\ Kolesik,
G.\ Korniss,
R.~A.\ Ramos, 
H.~L.\ Richards,
H.\ Tomita,
and
J. Vi\~{n}als
for helpful discussions.
S.W.S.\ and P.A.R.\ appreciate hospitality and support from the
Colorado Center for Chaos and Complexity
during the 1997 Workshop on Nucleation Theory and Phase Transitions.
This research was supported in part by the Florida State University
Center for Materials Research and Technology, by the FSU Supercomputer
Computations Research Institute, which is partially funded by the
U.~S.\ Department of Energy through Contract No.\ DE-FC05-85ER25000,
and by the National Science Foundation through Grants No.\ DMR-9315969,
DMR-9634873, DMR-9520325, and DMR-9871455.
Computing resources at the National Energy Research
Supercomputer Center were made available by the U.~S. Department of
Energy.
}



\begin{thebibliography}{10}
\bibitem{error_md}
This elementary relation was regrettably misprinted in
  Ref.~\protect\cite{side98-SD}  .
\end{thebibliography}


\clearpage
\widetext

 \begin{table}
 \begin{center}
 \caption{
  \label{table_constants_md}
  Parameters and constants used in this work.
  The values of the parameters $H_{0}$, $L$, and $T$ have
  been selected
  such that switching occurs via the multi-droplet mechanism.
  The constants $\Xi_{0}(T)$ and $K$
  are calculated from droplet theory 
  \protect\cite{lang6769,gnw80,ccga93,ccga94}
  for two-dimensional Ising systems.
  The constants
  $\left < \tau \right >$ and $r$ are measured from 
  field-reversal MC simulations
  with the Glauber dynamic
  (using the parameters listed in the left column).
  The constants
  $\Omega(T)$ \protect\cite{ccga94}
  and $\nu(T)$ \protect\cite{ramo97},
  where the droplet interface velocity is $v_{0} \! = \! \nu(T)|H|$,
  have been measured in other work.
  The value for $H_{\rm DSP}$ is taken from
  Fig. 11 of Ref.~\protect\cite{jlee95}.
  For $L \! = \! 90$ and $128$, the relative standard deviation is 
  $r \! = \! 0.072$
  and $0.053$ respectively.
  Except for $H_{\rm DSP}$, which decreases slowly with increasing $L$, 
  all other values are the same for $L \! = \! 90$
  and $L \! = \! 128$.
}
 \begin{tabular}{|| l | r || l | r || l | r ||}
  \multicolumn{2}{||c||}{Parameters} 
  & \multicolumn{2}{c||}{Constants (theory)} & 
  \multicolumn{2}{c||}{Constants (simulation)} \\
  \hline
  $H_{0}$  &   $0.3J$    &  $\Xi_{0}(T)$  & $0.506192 \ J$
                                                            
&   $\Omega_{2}(T)$   &  $3.15255$  \\
  \hline
  $L$      &   $64$       &  $K$        &     $3$ (exact)   &   $\nu(T)$
           & $(0.465 \pm 0.014) \ J^{-1}{\rm MCSS}^{-1}$ \\
  \hline
  $T$      &   $0.8T_{c}$     & \multicolumn{2}{c||}{}
           &   $\left < \tau \right > $     &  $74.5977$ MCSS \\
  \hline
  \multicolumn{2}{||c||}{} & \multicolumn{2}{c||}{}          
  & $H_{\rm DSP} (L \! = \! 64)$ & $(0.11 \pm 0.005) \ J$  \\
  \hline
  \multicolumn{2}{||c||}{}& \multicolumn{2}{c||}{}          
  & $r (L \! = \! 64)$          & $0.105$   \\
 \end{tabular}
 \end{center}
 \end{table}


\clearpage

\narrowtext
\clearpage

\clearpage

\pagebreak

\pagebreak

\pagebreak

\pagebreak

\end{document}